\documentclass[12pt]{article}
\usepackage{amsmath,amssymb} 
\usepackage{graphicx}
\usepackage{natbib}
\usepackage{url} 

\newcommand{\blind}{0}

\usepackage{stmaryrd} 
\usepackage{xcolor} 
\usepackage{nicefrac}
\usepackage[ruled,vlined,linesnumbered]{algorithm2e} 
\usepackage{algpseudocode} 
\usepackage{subcaption} 
\usepackage{booktabs} 
\usepackage{multirow} 

\usepackage{svg}

\newtheorem{theorem}{Theorem} 
\newtheorem{corollary}{Corollary} 

\newcommand{\norm}[1]{\left\lVert #1 \right\rVert}

\begin{document}

\def\spacingset#1{\renewcommand{\baselinestretch}%
{#1}\small\normalsize} \spacingset{1}


\if0\blind
{
  \title{\bf Clustering-Based Inter-Regional Correlation Estimation}
  \author{Hanâ Lbath$^{a,}$\thanks{Corresponding author: hana.lbath@inria.fr}
  , Alexander Petersen$^b$, Wendy Meiring$^c$, Sophie Achard$^{a}$\\ \hspace{.2cm}\\
   $^a$Univ. Grenoble Alpes, CNRS, Inria, Grenoble-INP, LJK, \\ 38000 Grenoble, France \vspace{.2cm} \\
    $^b$Department of Statistics, Brigham Young University, \\ Provo, UT 84602, USA \vspace{.2cm} \\
    $^c$Department of Statistics and Applied Probability,\\
    University of California Santa Barbara, Santa Barbara, CA 93106, USA \\
    } 
  \maketitle
  \vspace{-1cm}
} \fi

\if1\blind
{
  \bigskip
  \begin{center}
    {\LARGE\bf Clustering-Based Inter-regional Correlation Estimation}
\end{center}
  \medskip
} \fi

\begin{abstract}
A novel non-parametric estimator of the correlation between grouped measurements of a quantity is proposed in the presence of noise. This work is primarily motivated by functional brain network construction from fMRI data, where brain regions correspond to groups of spatial units, and correlation between region pairs defines the network. The challenge resides in the fact that both noise and intra-regional correlation lead to inconsistent inter-regional correlation estimation using classical approaches. While some existing methods handle either one of these issues, no non-parametric approaches tackle both simultaneously. To address this problem, we propose a trade-off between two procedures: correlating regional averages, which is not robust to intra-regional correlation; and averaging pairwise inter-regional correlations, which is not robust to noise. To that end, we project the data onto a space where Euclidean distance is used as a proxy for sample correlation. We then propose to leverage hierarchical clustering to gather together highly correlated variables within each region prior to inter-regional correlation estimation. We provide consistency results, and empirically show our approach surpasses several other popular methods in terms of quality. We also provide illustrations on real-world datasets that further demonstrate its effectiveness. \end{abstract}

\noindent%
\vfill

\newpage
\spacingset{1.5} 
\section{Introduction}

Correlation estimation is integral to a wide range of applications, and is often the starting point of further analyses. However, data are often contaminated by noise. If data are additionally inherently divided into separate, and study-relevant groups, inter-group correlation estimation becomes all the more challenging.
Such datasets are often encountered in spatio-temporal 
studies, such as single-subject brain functional connectivity network estimation, where voxel-level signals acquired via functional Magnetic Resonance Imaging (fMRI) are grouped into predefined spatial brain regions \citep{Fallani2014}. This work is relevant as well to other fields, such as organizational studies, where individuals are grouped by organization \citep{Ostroff19993aggregated}. 
As such, we will be using the words group, region, and parcellation interchangeably. In these contexts, measurement replicates of each individual element, 
most often collected across time, are available and used to compute the sample correlation between different regions. These elements are grouped according to a parcellation which is fixed and corresponds to a practical reality, like anatomical brain regions in fMRI studies. As a result, regions could themselves be inhomogeneous. 
This work hence aims to estimate inter-regional correlation, later shortened to inter-correlation, no matter the quality of the parcellation. However, both noise and arbitrary within-region 
correlation, later called intra-correlation, lead to inconsistent inter-correlation estimation by Pearson's correlation coefficient \citep{Ostroff19993aggregated,Saccenti2020CorruptionOT}. Indeed, it has been established in various contexts that correlation is underestimated in the presence of noise \citep{Ostroff19993aggregated,2017MatzkeBayesCorrError,Saccenti2020CorruptionOT}. Furthermore, data are often high dimensional, which presents a challenge of its own. In practice, including many fMRI studies, 
variables hence are commonly spatially averaged by regions prior to inter-correlation estimation  \citep{achard.2006.1,Fallani2014}. Yet, intra-correlation may be weak, which would lead to overestimation of inter-correlations \citep{1984wigleyAvgtimeseries}. This phenomenon may also be compounded by unequal region sizes \citep{Achard2011fMRIFunctionalConnectivity}. Thus, standard correlation estimators are not well-suited for 
the setting of grouped variables under noise contamination.
Nonetheless, simultaneously tackling noise and intra-group 
dependence structures can be quite difficult, especially in a non-parametric setting.
Failing to do so can be especially problematic for downstream analyses. For instance, in functional connectivity network estimation, a threshold is often applied to sample
inter-correlation coefficients in order to identify 
edges between brain regions. Under- or over-estimation of the inter-correlation would then lead to missing or falsely detecting edges. 

To address these problems, we present a data-driven, and non-parametric, approach with an astute intermediate aggregation. First, we propose to gather together highly correlated variables within each region. To this end, variables are projected onto a space where Euclidean distance can serve as a substitute to 
sample correlation, with lower values of the former corresponding to higher correlations. Hierarchical clustering with Ward's linkage \citep{Ward1963,Murtagh2014hclust} is then applied to the projected variables within each region, resulting in intra-regional clusters of highly correlated variables. Within each intra-regional cluster, these variables are next spatially averaged. For each pair of regions, a sample correlation is then computed for each pair of cluster-averages from different regions. Our approach hence provides a distribution of the sample 
inter-correlations between each pair of regions, containing as many sample correlations as there are pairs of clusters from the two regions. For a point estimate of the inter-regional correlation for a given pair of regions, the average of the sample inter-correlation coefficients can then be considered. We summarize our main contributions as follows:
\begin{itemize}
    \item We propose a novel non-parametric estimator of inter-regional
    correlation that offsets the combined effect of noise and arbitrary intra-correlation 
    by leveraging hierarchical clustering. 
    \item Based on the properties of hierarchical clustering with Ward's linkage, we prove our estimator is consistent for an appropriate choice of the cut-off height of the dendrograms thus obtained.
    \item We then empirically corroborate our results about the impact of the cut-off height on the quality of the estimation. We also show our proposed inter-correlation estimator outperforms popular estimators in terms of quality, and illustrate its effectiveness on real brain imaging datasets.
\end{itemize}

\section{Related Work}
\label{sec:biblio}

In the context of functional connectivity, the vast majority of papers that build correlation networks first average signals within each brain region for each time point, before computing Pearson's correlation across time, possibly after wavelet or other filtering, e.g., \citep{achard.2006.1,Bolt2017CA,Akitoshi2021CA,Zhang2016CA}. Nevertheless, and as mentioned in the previous section, the correlation of averages overestimates the true correlation when intra-regional correlations are weak, while high noise may lead to underestimation. It was also empirically observed in fMRI data that the application of spatial smoothing, which is a common preprocessing step 
to reduce the effect of noise, causes the inter-regional correlations to be overestimated \citep{LiuSmoothingFmri}.

Several methods tackling the impact of intra-correlation 
on the estimation of inter-correlation have been proposed in familial data literature, e.g.,\citep{Elston1975Cor,rosner.1977.1,Srivastava1988intercorr,2010wilson_families}. These approaches nonetheless do not address the impact of noise. Moreover, they require normality assumptions on the samples, while we provide consistency guarantees for our proposed estimator that do not require parametric assumptions on the signal distribution. Bayesian inference methods have been proposed to offset the effect of measurement errors \citep{2017MatzkeBayesCorrError}. However they require a careful choice of priors, in addition to only handling pairs of variables, as opposed to 
groups of variables---which is what we are interested in. Robust correlation estimation has also been extensively investigated but mostly for specific distributions, such as contaminated normal distributions \citep{Shevlyakov_Smirnov_2016} or with heavy tails \citep{lindskog2000linear}, whereas we are interested in robustness to noise and weak intra-group dependence. Furthermore, groups of variables are not considered either. Cluster-robust inference in the presence of both noise and within-group 
correlation has been studied in the econometric literature \citep{ColinCameron2015APG}. However, inter-correlation, which is the quantity we aim to estimate in this work, is assumed to be zero. To the best of our knowledge, we are the first to propose a method to simultaneously tackle the impact of noise and within-group inhomogeneity to estimate inter-correlation in a non-parametric fashion.

\section{Preliminaries}
\label{sec:prelim}

From this point forward, and without loss of generality, we will focus on spatio-temporal contexts. In particular, we are motivated by an application to brain fMRI data where individual observed variables correspond to blood-oxygen-level-dependent (BOLD) signals that are assigned to \textit{voxels}, and are grouped by \textit{regions}. Nonetheless, the following results can be applied to any dataset of grouped measurements of a quantity. In this section we define our notation and model, together with the inter- and intra-correlation coefficients. Throughout this paper we consider two regions, generically denoted $A$ and $B$. In reality, datasets will involve a potentially large number of regions but, for the purpose of correlation network construction, the correlations can be estimated in a pairwise fashion at the regional-level.
Let $X^A_1, \dots, X^A_i,  \dots, X^A_{N_A}$ denote $N_A$ spatially dependent latent (unobserved) random variables in region $A$, each variable corresponding to an individual voxel in that region.
   Let $\epsilon^A_1, \dots, \epsilon^A_i,  \dots, \epsilon^A_{N_A}$ represent 
   random noise variables. 
    We assume that the latent process $X_i^A$ at each voxel $i$ is contaminated by noise $\epsilon^A_i$, so that the observed variables $Y^A_i$ in region $A$ are
\begin{equation}
    \label{eq:model}
    Y^A_i = X^A_i + \epsilon_i^A, \quad i = 1,\ldots,N_A.
\end{equation}
    We assume within-region homoscedasticity of both signal and noise, i.e., 
\begin{equation*}
     \sigma_A^2 = {\rm Var}\left(X_i^A\right), \,\,
     \gamma_A^2
     = {\rm Var}\left(\epsilon_i^A\right), \quad \ \ i = 1, \ldots,  N_A.
\end{equation*}
    Analogously we define $N_B$, $X_j^B$, $\epsilon_j^B$, $Y_j^B$, $\sigma_B^2$ and 
    $\gamma_B^2$,
for region $B$ and voxels $j = 1,\ldots,N_B$. We assume the noise variables are spatially uncorrelated both within and across regions, and that they are also uncorrelated to the latent state both within and between regions. A critical reality of the observed data is the \textit{intra-correlation} or Pearson's correlation between any pair 
of 
random variables \textit{within} a given region $A$.  We denote by $\eta_{i,i^\prime}^{A}$ the intra-correlation of the latent variables $X_i^A, X_{i^\prime}^A$. We place no further constraints on the intra-correlation structure. Similarly, we define the 
\textit{inter-correlation} as Pearson's correlation between any pair of random variables from two \textit{distinct} regions. For a given pair of distinct regions, $A, B$, the inter-correlation between any pair of latent random variables $X_i^A, X_j^B$ is assumed to be constant across voxels, and is denoted as $\rho^{A,B}$.

Consider now $n$ temporally independent and identically distributed (i.i.d.) 
samples of all observed signals. That is, for each region $A$ and voxel $i = 1, \dots, N_A$, we have $n$ i.i.d. observations $Y_i^A(t),$ $t = 1,\ldots,n,$ each distributed as in \eqref{eq:model} with the same intra- and inter-correlation properties as those outlined previously.  In particular, for any time point
$t = 1, \dots,  n $, and voxels $i$ and $j$ from distinct regions $A$ and $B$, respectively, $Cov(Y^A_i(t), Y^B_j(t) ) =\rho^{A,B} \sigma_A \sigma_B$. Denote by $\textbf{Y}_i^A = [Y_i^A(1), \dots, Y_i^A(t), \dots Y_i^A(n)]$ the vector of observations for the $i$-th voxel of region $A$.

\section{Proposed Inter-Correlation Estimator}
\label{sec:estimator}

After defining the 
sample correlation coefficient in Section \ref{samplecor}, we highlight in Section \ref{biases} the impact of the combined presence of noise and intra-correlation, when using popular 
estimators of inter-correlation. In Section \ref{proposedEstimator} we then propose an inter-correlation estimator that limits these effects. Consistency of our estimator is proved in Section \ref{sec:consistency}. 

\subsection{Computing Sample Correlations}
\label{samplecor}
We denote by $\widehat{Cor}(\cdot , \cdot)$ the sample (Pearson's) correlation between any two equal-length vectors of samples. This corresponds to the zero-lag empirical cross-correlation in spatio-temporal studies. To be specific, suppose $\mathbf{a}, \mathbf{b} \in \mathbb{R}^n$ are any vectors of the same length, and let $\overline{a} = n^{-1}\sum_{t = 1}^n a_t$ and $\overline{b}=n^{-1}\sum_{t = 1}^n b_t$ be the averages of their elements, respectively.  Let $\mathbf{1}_n$ be the $n$-vector of ones, $\mathbf{a}^c = \mathbf{a} - \overline{a}\mathbf{1}_n,$ and $\mathbf{b}^c = \mathbf{b} - \overline{b}\mathbf{1}_n$ their centered versions. With $\langle \cdot, \cdot \rangle$ and $\norm{\cdot}$ being the Euclidean inner product and norm, respectively, we define
\begin{equation}
    \label{eq:corHatDef}
    \widehat{Cov}(\mathbf{a},\mathbf{b}) = n^{-1}\langle \mathbf{a}^c,\mathbf{b}^c\rangle, \,\, \widehat{Var}(\mathbf{a}) = n^{-1}\norm{\mathbf{a}^c}^2, \,\,
    \widehat{Cor}(\mathbf{a},\mathbf{b}) = \frac{\widehat{Cov}(\mathbf{a}, \mathbf{b})}{\sqrt{\widehat{Var}(\mathbf{a})\widehat{Var}(\mathbf{b})}}.
\end{equation}

Using this notation, the sample correlation between any two voxels $i$ and $j$ in regions $A$ and $B$ is
 	\begin{equation}
	 \label{eq:samplecor}
	    	r_{i,j}^{A,B} = \widehat{Cor}(\textbf{Y}_{i}^{A}, \textbf{Y}_{j}^{B}). 
	\end{equation}	
Observe that this definition applies equally to 
sample inter-correlations ($A \neq B$) as well as intra-correlations ($A = B$).

\subsection{Impact of Noise and Intra-Correlation}
\label{biases}
Previously, \cite{2017MatzkeBayesCorrError} showed that the presence of noise attenuates the observed correlation. Indeed, this phenomenon is captured in the following result: from model \eqref{eq:model} and  \cite{Achard2020RobustCF}, $r_{i,j}^{A,B}$ converges almost surely to 
\begin{align}
    \label{eq:vox}
    \frac{ Cov(Y_i^A, Y_j^B) }{ \sqrt{(\sigma_A^2 + \gamma_A^2)\cdot(\sigma_B^2 + \gamma_B^2) } }
    =  \frac{ Cov(X_i^A, X_j^B) }{ \sqrt{(\sigma_A^2 + \gamma_A^2)\cdot(\sigma_B^2 + \gamma_B^2) } }. 
\end{align}
Therefore, if distinct regions $A$,$B$ with latent signals observed contaminated by noise, $r_{i,j}^{A,B}$ is not a consistent estimator of true inter-correlation $\rho^{A,B}$ due to the presence of the noise variances in the denominator of \eqref{eq:vox}. Furthermore, in settings where a single point estimate of the inter-correlation of the unobserved latent signal between two regions is needed, the corresponding pairwise sample 
inter-correlation coefficients can be averaged to provide an estimator. Denoted $r^{AC}_{A,B}$, it corresponds to the ensemble estimator in 
familial data literature 
\citep{rosner.1977.1}:
\begin{equation}
    r^{AC}_{A,B} = \frac{1}{N_A \cdot N_B} \sum\limits_{i=1}^{N_A} \sum\limits_{j=1}^{N_B} r_{i,j}^{A,B}. 
\end{equation}
However, the latter is similarly impacted by noise.

As mentioned in Section \ref{sec:biblio}, one of the most popular estimators in neuroimaging studies consists of spatially averaging the observation random variables within each distinct region for each time $t$, before computing the 
sample correlation between these averages. Specifically, define regional (spatial) averages $\overline{\textbf{Y}}^A = N_A^{-1}\sum_{i = 1}^{N_A} \textbf{Y}_i^A$ and $\overline{\textbf{Y}}^B = N_B^{-1}\sum_{j = 1}^{N_B} \textbf{Y}_j^B$.  
Then this estimator is
    \begin{equation}
         r^{CA}_{A,B} = \widehat{Cor}(\, \overline{\textbf{Y}}^A, \overline{\textbf{Y}}^B\, ).
         \label{eq:corrAve}
    \end{equation}
Under model \eqref{eq:model}, and according to results from \citep{Achard2020RobustCF}, together with intra-regional uncorrelatedness between latent and noise random variables, as well as inter-regional uncorrelatedness of noise, $r^{CA}_{A,B}$ converges almost surely to: 
\begin{align}
    \label{eq:rCA}
    \frac{\rho^{A,B} 
    }{ \sqrt{ \left[ \frac{1}{N_A^2} \cdot \sum\limits_{i, i^\prime=1}^{N_A} \eta_{i,i^\prime}^{A} + \frac{\gamma_A^2}{N_A \cdot \sigma_A^2} \right] \left[ \frac{1}{N_B^2} \cdot \sum\limits_{j, j^\prime=1}^{N_B} \eta_{j,j^\prime}^{B} + \frac{\gamma_B^2}{N_B \cdot \sigma_B^2}  \right]} },
\end{align} 
where $N_A^{-2} \cdot \sum_{i,i^\prime=1}^{N_A} \eta_{i,i^\prime}^{A}$ is the spatial average of the pairwise latent intra-correlation coefficients 
within region $A$. 

It follows from \eqref{eq:rCA} that intra-correlation and noise both contribute to inconsistency of the 
inter-correlation estimator \eqref{eq:corrAve}. 
Indeed, both quantities appear in the denominator. It is then apparent that the smaller the regions (smaller $N_A$), the higher the impact of noise on the correlation estimation. Additionally, the weaker the spatial intra-regional dependence, the larger the overestimation of the inter-correlation. This effect may also be compounded when regions are large, as was observed by \cite{Achard2011fMRIFunctionalConnectivity}. One would then need to have regions as large 
as possible, while having an average intra-correlation as close to 1 as possible in order to offset these biases. However, large regions tend to be inhomogeneous in practical scenarios, and thus tend to have low intra-correlation. 

\subsection{A Clustering-Based Inter-Correlation Estimator}
\label{proposedEstimator}

Based on these findings, we propose an  inter-correlation estimator specifically designed to limit the combined effects of noise and intra-correlation. Instead of aggregating over entire regions, we propose to aggregate over small groups of highly intra-correlated variables (cf. Steps 1 and 2), before computing the correlation of the corresponding local averages (cf. Step 3). 

\subsubsection{Step 1: U-Scores Computation}
\label{sec:uscore}
To facilitate the grouping of the variables within each region, we can leverage U-scores to
project the sample vectors $\textbf{Y}^A_i$ onto a space where the Euclidean distance can be used as a proxy for the 
sample correlations.We could then apply any clustering algorithm in the U-score space. \textit{U-scores} are an orthogonal projection of the Z-scores of random variables onto a unit $(n-2)$-sphere centered around 0. The U-score $\textbf{U}_i^{A}$ of $\textbf{Y}_i^{A}$ is defined by $\textbf{U}_i^{A} = \textbf{H}_{2:n}^T \textbf{Z}_i^{A}$, where $\textbf{H}_{2:n}^T$ is a $(n-1) \times (n-1)$ matrix obtained by Gramm-Schmidt orthogonalization, and $\textbf{Z}_i^{A}$ the Z-score of $\textbf{Y}_i^{A}$. We refer to \citep{hero2011} for a full definition. Sample correlations can then be expressed as an inner product of U-scores: $r_{i,j}^{A,B} = (\textbf{U}_i^A)^T \textbf{U}_j^B = 1 -  \| \textbf{U}_i^A - \textbf{U}_j^B \|^2 /2$, where  $\textbf{U}_i^A$, $\textbf{U}_j^B$ are the U-scores of the $i$th and $j$th voxels 
in regions $A$ and $B$, respectively, and $\|\cdot\|^2$ is the squared Euclidean distance. 

\subsubsection{Step 2: Clustering}
\label{sec:clustering}
Once the U-scores are calculated, any standard clustering algorithm can be applied to obtain homogeneous groups of variables within each region. Agglomerative hierarchical clustering with Ward's linkage  \citep{Ward1963,Murtagh2014hclust}, which is closely related to the k-means algorithm \citep{kmeans}, aims to minimize the intra-cluster variance, which implies a maximization of the intra-cluster correlation. A comparison of different clustering methods, which empirically validates the use of Ward's linkage in our context, is presented in Section \ref{sec:choiceMethod}. In practice, the number of clusters generally needs to be specified. However, such a strategy, 
while often satisfactory in common clustering tasks, such as exploratory analyses, does not provide any obvious theoretical guarantees on the homogeneity of the clusters, which is what we are interested in. Nevertheless, hierarchical clustering outputs a dendrogram that can then 
be cut off at a designated height to produce 
a clustering. Therefore, instead of setting a number of clusters, we propose to specify a cut-off height through which cluster radii, and by proxy intra-correlations, can be controlled to a certain extent (cf Theorem \ref{th:ineq.h}). Proofs can be found in the appendix.

\begin{theorem}
\label{th:ineq.h}
For a region $A$, a fixed cut-off height $h_A$, and all clusters $\nu_A$ thus obtained, the spatial average of the sample 
intra-cluster correlation is bounded as follows:
\begin{align}
    \label{eq:ineq.h}
    1 - \frac{h_A^2}{2} \ \leq \ 
    \frac{1}{|\nu_A|^2} \sum\limits_{i, i^{\prime}=1}^{|\nu_A|} r_{i, i^{\prime}}^{A,A}
    \ 
    \leq 1, 
\end{align}
where $|\nu_A|$ is the size of cluster $\nu_A$.
\end{theorem}

%
Theorem \ref{th:ineq.h} shows that through careful choice of the cut-off heights, clusters of highly correlated variables can be selected within each region. This choice can be guided by the ensuing 
observations about the maximum distance between U-scores within a given region, denoted by $h_A^{\text{max}}$, which follow immediately from Theorem \ref{th:ineq.h} and the fact that $1 - ({h_A^{\text{max}}})^2/2 \ = \min\limits_{i,i^{\prime} = 1,\dots,N_A} r_{i,i^{\prime}}^{A,A}$: 
\begin{itemize}
    \item if $h_A \geq h_A^{\text{max}}$, 
\begin{equation}
    \label{eq:h.geq.hmax}
    1 - \frac{h_A^2}{2} \ \leq \ \min\limits_{i,i^{\prime}=1,\dots,N_A} r_{i,i^{\prime}}^{A,A}  \ \leq \  
     \frac{1}{|\nu_A|^2} \sum\limits_{i, i^{\prime}=1}^{|\nu_A|} r_{i, i^{\prime}}^{A,A}
\end{equation}

    \item and if $h_A \leq h_A^{\text{max}}$, 
\begin{equation}
    \label{eq:h.leq.hmax}
    \min\limits_{i,i^{\prime}=1,\dots,N_A} r_{i,i^{\prime}}^{A,A} \ \leq \ 1 - \frac{h_A^2}{2} \ \leq \  
     \frac{1}{|\nu_A|^2} \sum\limits_{i, i^{\prime}=1}^{|\nu_A|} r_{i, i^{\prime}}^{A,A}.
\end{equation}
\end{itemize}
Therefore, to ensure all clusters contain more than one voxel, the maximum distance between any two clusters of the region (i.e., the cut-off height) would need to be larger than the maximum distance between any two voxels within the region (i.e., $h_A^{\text{max}}$).
Thus, setting the cut-off height to $h_A^{\text{max}}$ would ensure to obtain the smallest possible clusters guaranteed to contain at least two variables.
Moreover, computing $h_A^{\text{max}}$ is computationally inexpensive. It also does not depend on any ground-truth, which remains unknown in practice. 
Empirical explorations of an optimal choice are presented in Section \ref{sec:choiceH}, and demonstrate the practical effectiveness of setting the cut-off height to $h_A^{\text{max}}$.

\begin{figure}[t!]
    \centering
    \includegraphics[width=\linewidth]{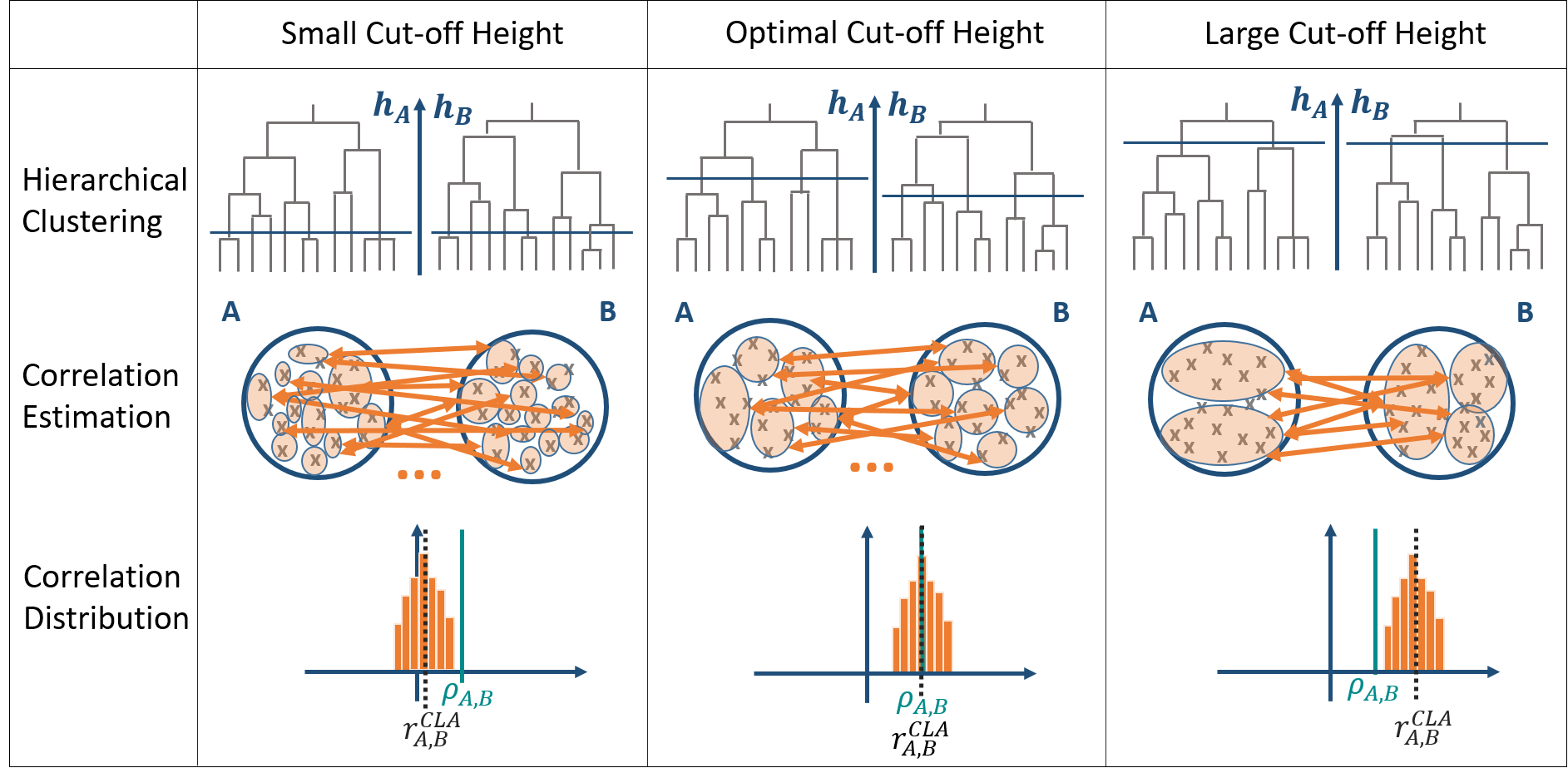}
    \caption{Illustration of the inter-correlation estimation 
    of a pair of regions for different cut-off heights. The top panel shows the dendrograms of the hierarchical clustering applied to each region. The horizontal line over each 
    dendrogram indicates the cut-off heights $h_A, h_B$. The grey crosses in the middle panel correspond to the random variables inside each regions, and are grouped into the resulting clusters (orange ellipses). 
    The arrows represent the sample 
    inter-correlation between the average of the variables inside each cluster (some arrows were left out to improve readability). The bottom panel displays the distribution of the pairwise sample 
    inter-correlation. The true inter-correlation $\rho_{A,B}$ (solid line) is best approximated by the sample 
    inter-correlation $r_{A,B}^{CLA}$ (dotted line) when the cut-off heights are neither too small nor too large.  
    } 
    \label{fig:clustCLA}
\end{figure}

\subsubsection{Step 3: Clustered Correlation Estimation} Once clusters are obtained within each region, the inter-correlation is estimated as follows. For two distinct regions $A$ and $B$, for fixed cut-off heights $h_A, h_B$, and any two pairs of clusters $\nu_A, \nu_B$ within each of these regions, we define the following cluster-level inter-correlation estimator: 
\begin{equation}
    \label{eq:rCLAdef}
    r^{CLA}_{\nu_A, \nu_B} = \widehat{Cor}(\ \overline{\textbf{Y}}^{\nu_A}, \  \overline{\textbf{Y}}^{\nu_B}\, ), 
\end{equation}
where  $\overline{\textbf{Y}}^{\nu_A} = |\nu_A|^{-1}\sum_{i \in \nu_A} \textbf{Y}_i^A$, and $\overline{\textbf{Y}}^{\nu_B}$ is defined similarly. A distribution of sample inter-correlation coefficients is hence obtained for this 
pair of regions, as seen in Figure \ref{fig:clustCLA}. As mentioned earlier, if a point estimate is needed, one can then simply average the cluster-level estimates to derive the following regional-level estimator:
\begin{align}
    \label{eq:rCLA_AB}
    r^{CLA}_{A,B} = \frac{1}{ N^{clust}_A \cdot N^{clust}_B } \sum_{\nu_A, \nu_B} r^{CLA}_{\nu_A, \nu_B},
\end{align}
where $N^{clust}_A$ is the number of clusters within region $A$. We refer to Algorithm \ref{algo:clustcorr} for a detailed description of our proposed clustering-based correlation estimation procedure for $J$ regions.

\begin{algorithm}[h!]
     \SetKwData{Left}{left}\SetKwData{This}{this}\SetKwData{Up}{up}
    \SetKwFunction{Union}{Union}\SetKwFunction{FindCompress}{FindCompress}
    \SetKwInOut{Input}{input}\SetKwInOut{Output}{output}
    
    \Input{$N$ variables grouped in $J$ regions with $n$ samples each}
    \Output{
    Cluster-level and regional-level inter-correlation estimates}
    \algorithmiccomment{Clustering}
   
    \For{each region $A$}{
        Apply hierarchical clustering to $A$;
        
        Choose the cut-off height $h_A$;
        
        Cut the dendrogram at height $h_A$;
        
        \For{each cluster $\nu_A$ in $A$}{
        $\overline{\textbf{Y}}^{\nu_A} \leftarrow \sum_{i=1}^{|\nu_A|} \textbf{Y}^A_i / |\nu_A|$; 
        
        }
    }
    
    \Comment{Correlation of local averages estimation}
    
    \For{each pair of regions $A,B$}{
        \For{each pair of clusters $\nu_A, \nu_B$}{
            $r^{CLA}_{\nu_A, \nu_B} \leftarrow \widehat{Cor}(\overline{\textbf{Y}}^{\nu_A}, \overline{\textbf{Y}}^{\nu_B}) $ 
        }
        
        $r^{CLA}_{A, B} \leftarrow \sum_{\nu_A, \nu_B} r^{CLA}_{\nu_A, \nu_B}/ N^{clust}_A \cdot N^{clust}_B $
    }
    
    \caption{Clustering-Based Correlation Estimation}
 \label{algo:clustcorr}
\end{algorithm}


\subsection{Consistency of the Proposed Estimator}
\label{sec:consistency}
The clusters derived in Algorithm \ref{algo:clustcorr} are data-driven, and thus random from a probabilistic perspective. To simplify analysis and allow us to demonstrate the expected behavior of the proposed estimator as the 
number of time points $n$ grows, let us assume that clusters $\nu_A$ and $\nu_B$ are fixed. Then define the following quantity, which will be used in several of the subsequent results:
\begin{align}
    \rho^{CLA}_{\nu_A, \nu_B} = \frac{ \rho^{A,B}}{ \sqrt{ \left[ \frac{1}{|\nu_A|^2} \cdot \sum\limits_{i,i^\prime=1}^{|\nu_A|} \eta_{i,i^\prime}^{A} + \frac{\gamma_A^2}{|\nu_A| \cdot \sigma_A^2 }  \right] \cdot \left[ \frac{1}{|\nu_B|^2} \cdot \sum\limits_{j,j^\prime=1}^{|\nu_B|} \eta_{j,j^\prime}^{B} + \frac{\gamma_B^2}{|\nu_B| \cdot \sigma_B^2  }  \right]} }.
\end{align}
\begin{theorem}
\label{th:consistent}
Under the assumptions of model \eqref{eq:model}, for a 
fixed pair of clusters $\nu_A, \nu_B$, as $n$ tends towards infinity, 
    \begin{align}
        \label{eq:consistency}
        r^{CLA}_{\nu_A, \nu_B} & \  \stackrel{a.s.}{\to} 
        \ 
        \rho^{CLA}_{\nu_A, \nu_B} . 
    \end{align}
\end{theorem}


The proof is detailed in the appendix. We obtain similar results for the regional-level point estimate $r^{CLA}_{A,B}$.

\begin{corollary}
    \label{cor:consistency}
    Under the same assumptions as Theorem \ref{th:consistent}, for two regions $A, B$, as $n$ tends towards infinity, 
\begin{align}
    \label{eq:consistency_AB}
    r^{CLA}_{A, B} \ \stackrel{a.s.}{\to} \ \frac{1}{N^A_{clust} N^B_{clust}}  \sum\limits_{\nu_A, \nu_B}  
    \rho^{CLA}_{\nu_A, \nu_B}.
\end{align}
\end{corollary}
Corollary \ref{cor:consistency} is a direct consequence of Theorem \ref{th:consistent}. 

Theorem \ref{th:consistent} and Corollary \ref{cor:consistency} emphasize the fact that controlling the denominator of $\rho_{\nu_A,\nu_B}^{CLA}$ is key to obtaining a consistent estimator of $\rho^{A,B}$. This brings to light 
the influence of the cut-off height, and thereby the cluster size and intra-cluster correlation, on the consistency of the inter-correlation estimate, both at the cluster- 
and regional-level.

For a pair of regions $A,B$, as the cut-off heights $h_A, h_B$ become larger, the impact of noise diminishes. Moreover, the clusters increase in size until there is only a single cluster left that corresponds to the entire region. Thus, for $h_A, h_B$ sufficiently large, our proposed estimator $r^{CLA}_{\nu_A,\nu_B}$, and the corresponding point estimate $r^{CLA}_{A,B}$ are equal to the correlation of averages $r^{CA}_{A,B}$ mentioned earlier. Conversely, as $h_A, h_B$ become smaller 
the maximum distance between U-scores within a cluster decreases, hence the minimal intra-cluster correlation increases (cf. Theorem \ref{th:ineq.h}). 
There are also gradually less variables within each cluster, until they eventually contain only a single variable. It follows that when $h_A, h_B = 0$, $r^{CLA}_{A,B}$ corresponds to a 
correlation estimate with no aggregation $r^{AC}_{A,B}$. This can be visualized in Figure \ref{fig:clustCLA}, where sample correlation distributions are depicted for different cut-off heights.
 
Therefore, to simultaneously lessen the impact of noise and intra-correlation a trade-off is necessary between a sufficiently high cut-off height (to decrease the impact of noise), and a low enough height (to decrease the impact of intra-cluster correlation). In such cases, both $r^{CLA}_{\nu_A, \nu_B}$ and $r^{CLA}_{A,B}$ are consistent estimators of the population inter-correlation.

\section{Experimental Results}
\label{sec:empirical}

In this section we empirically determine the optimal cut-off height, evaluate our proposed inter-correlation estimator on synthetic data, and illustrate our approach on real-world datasets.

\subsection{Datasets}
We first present the different datasets used in this paper.

\subsubsection{Real-World Datasets} 

\paragraph{Rat Brain fMRI Dataset}
We apply our estimator on fMRI data acquired on both dead and anesthetized rats \citep{becq_10.1088/1741-2552/ab9fec,guillaume2020functional}. In this paper we consider the following anesthetics: Etomidate (EtoL), Isoflurane (IsoW) and Urethane (UreL). The dataset is freely available at \url{https://dx.doi.org/10.5281/zenodo.7254133}. The scanning duration is $30$ min with a time	repetition of $0.5$ s. After preprocessing \citep{guillaume2020functional}, $25$ groups of voxels, each associated with its BOLD signal with a number of time points in the order of thousands, were extracted for each rat. They correspond to rat brain regions defined by an anatomical atlas obtained from a fusion of the Tohoku and Waxholm atlases \citep{guillaume2020functional}. Region sizes vary from about $40$ up to approximately $200$ voxels.  

\paragraph{Human Connectome Project} We also consider 
35 subjects from the human connectome project (HCP), WU-Minn Consortium pre-processed \citep{glasser_minimal_2013}. Subjects were pseudonymized. Two fMRI acquisitions on different days are available for each subject. The scanning duration is $14$ min and $24$ s with a time repetition of $720$ ms. A modified AAL template is used to parcellate the brain into $89$ regions. The details of the pre-processing are available in~\citep{termenon2016reliability}. Region sizes are in the order of thousands of voxels, and number of time points are in the order of thousands.

\subsubsection{Synthetic Datasets}
We consider several synthetic datasets to evaluate our estimator. For each simulation, we simultaneously generate $800$ independent samples of a pair of inter-correlated regions, containing each $60$ intra-correlated variables that follow a multivariate normal distribution with a predefined covariance structure contaminated by Gaussian noise. The inter-correlation is constant across all pairs of voxels. The different parameters are chosen 
to ensure the population covariance matrix of the two regions is positive semidefinite. For instance, one cannot generate a covariance matrix where both intra- and inter-correlation values are low. 

\begin{figure}[h!]
    \centering\begin{subfigure}{0.4\textwidth}
			\centering
			\includegraphics[width=.9\linewidth]{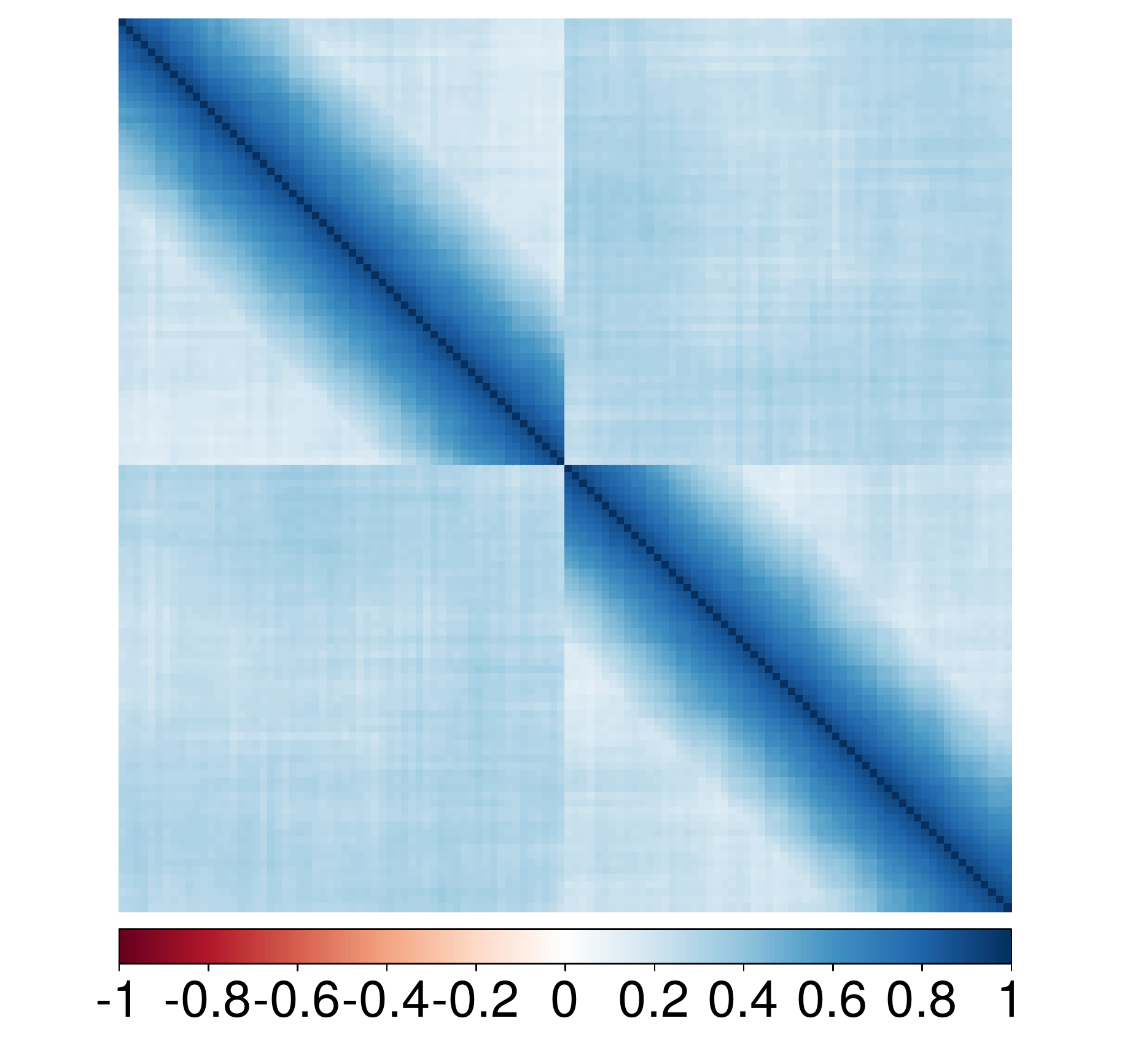}
			\caption{
			$\eta_A^{-}=\eta_B^{-}=0.2$}
		\end{subfigure}
		~
		\begin{subfigure}{0.4\textwidth}
			\centering
			\includegraphics[width=.9\linewidth]{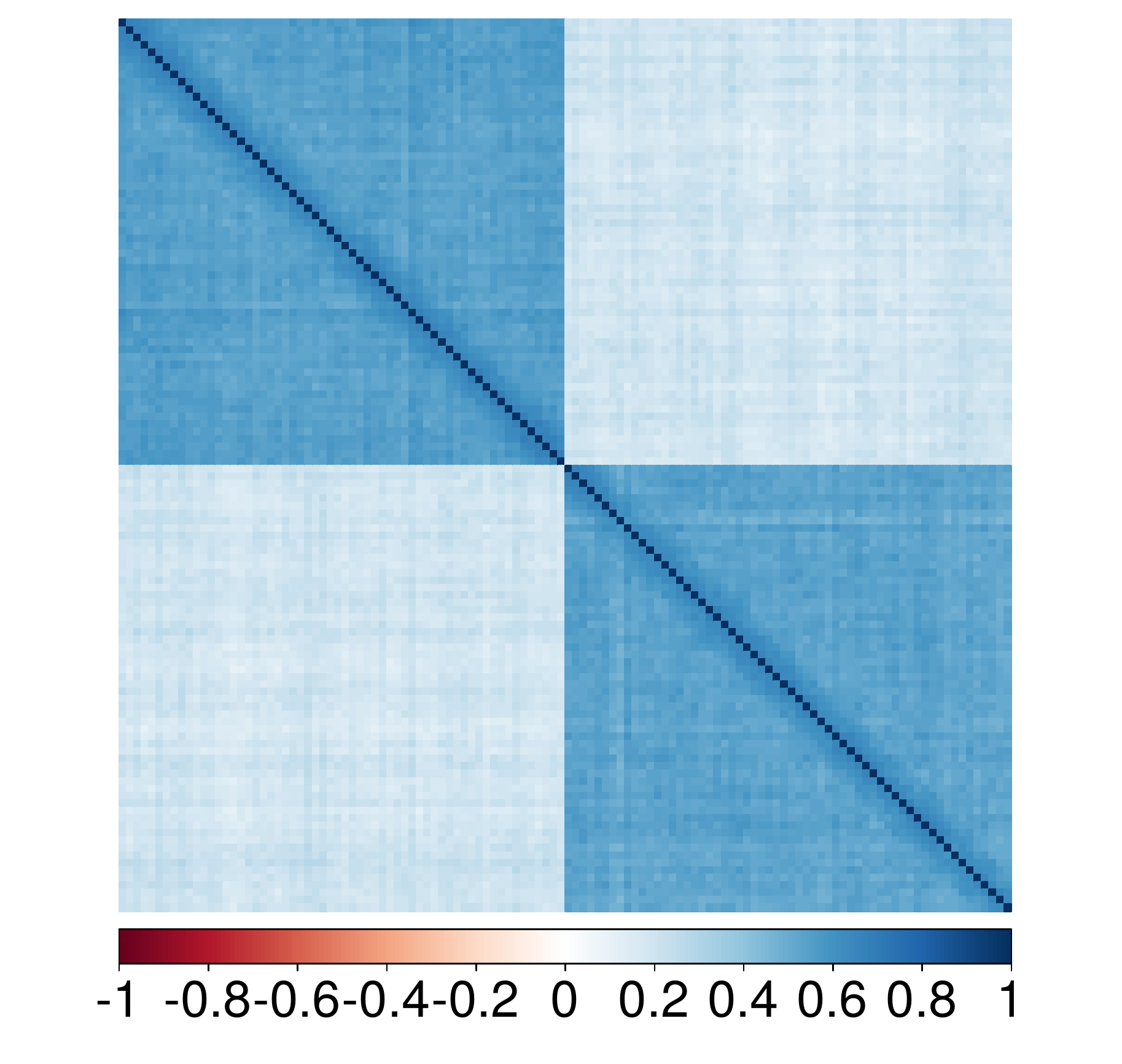}
			\caption{
			$\eta_A^{-}=\eta_B^{-}=0.8$}
		\end{subfigure}
    \caption{
    Sample pairwise correlation matrices (from the 1D Toeplitz model) for different minimum intra-correlation values, with an inter-correlation $\rho^{A,B}=0.3$ and noise variance $\gamma_A^2 = \gamma_B^2 = 0.5$. The diagonal blocks correspond to the intra-correlation of the two regions. }
    \label{fig:corrmat_simu}
\end{figure}

\paragraph{Toeplitz Covariance Structure}
We first generate 1-dimensional data with a Toeplitz intra-regional covariance structure (later denoted 1D Toeplitz). For each region, intra-correlation is defined such that it decreases as the distance between two variables increases: for any voxel $i, i^\prime$ in region $A$, $Cor(X_i^A, X_{i\prime}^A) = \max( 1 - |i^\prime-i|/30, \eta_A^{-} )$, where $|i^\prime-i|$ is the uniform norm between voxels $i$ and $i^\prime$, and  $\eta_A^{-}$ the minimal population intra-correlation of a region $A$. In this paper we consider several experimental settings by varying the population intra-correlation, inter-correlation and the variance of the noise. The sample pairwise correlation matrices of the 
observed signals are represented 
in Figure \ref{fig:corrmat_simu} for a low intra-correlation and a high intra-correlation setting with high noise.

\paragraph{Matérn Covariance Structure} Similarly we then simulate 3-dimensional data with a Matérn intra-regional covariance structure that depends on the Euclidean distance (later denoted 3D Matérn) \citep{geoR}. In this paper, we set the smoothness parameter to $\kappa_A = \kappa_B = 70$ to maintain the positive-definiteness of the input covariance matrix. 
We then vary the range parameters $\phi_A$, $\phi_B$ and the variance of the noise. The lower the range parameter, the lower the mean intra-correlation.

\paragraph{Spherical Covariance Structure} We then generate 3-dimensional data with a spherical intra-regional covariance structure that also depends on the Euclidean distance between voxels (later denoted 3D Spherical) \citep{geoR}. We vary the range parameters $\phi_A$, $\phi_B$ 
and the variance of the noise. The lower the range parameter, the lower the mean intra-correlation.

\subsection{Choice of the Cut-off Heights}
\label{sec:choiceH}
In this section we empirically evaluate on the 1D-Toeplitz dataset the impact of the  cut-off heights $h_A, h_B$ on the proposed clustering-based correlation estimator. We also propose a heuristic 
to choose optimal cut-off heights.

We consider different scenarios, including one that 
loosely matches live rat data settings, where the noise is high and the intra-correlation low.
For each simulated pair of regions, and for various cut-off heights $h_A, h_B$, the squared error of the cluster-level estimators are computed and then averaged across the different clusters: 
\begin{equation}
    \text{ERROR} = \frac{1}{N_{clust}^A N_{clust}^B} \sum\limits_{\nu_A, \nu_B} (r^{CLA}_{\nu_A, \nu_B} - \rho^{A,B} )^2.
\end{equation}

\begin{figure}[h!]
    \centering
    \begin{subfigure}{0.42\textwidth}
			\centering
            \includegraphics[width=\linewidth]{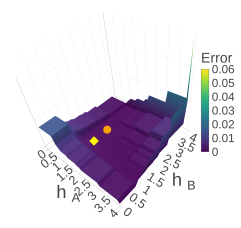}
			\caption{$\eta_A^{-}, \eta_B^{-}=0.2, \sigma_{\epsilon_A}^2, \sigma_{\epsilon_B}^2=0.5$}
		\end{subfigure}
		~
		\begin{subfigure}{0.42\textwidth}
			\includegraphics[width=\linewidth]{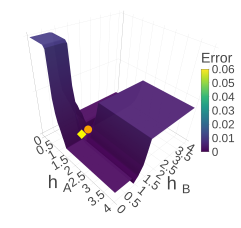}
			\caption{$\eta_A^{-}, \eta_B^{-}=0.8, \sigma_{\epsilon_A}^2, \sigma_{\epsilon_B}^2=0.5$}
		\end{subfigure}
		~
		\begin{subfigure}{0.42\textwidth}
			\centering
			\includegraphics[width=\linewidth]{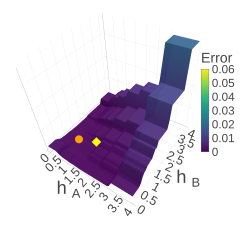}	
           \caption{$\eta_A^{-}, \eta_B^{-}=0.2, \sigma_{\epsilon_A}^2, \sigma_{\epsilon_B}^2=0.1$}
		\end{subfigure}
		~
		\begin{subfigure}{0.42\textwidth}
			\centering
			\includegraphics[width=\linewidth]{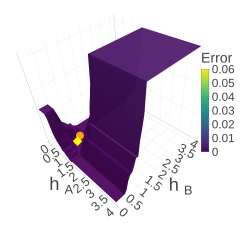}
            \caption{$\eta_A^{-}, \eta_B^{-}=0.8, \sigma_{\epsilon_A}^2, \sigma_{\epsilon_B}^2=0.1$}
		\end{subfigure}
    \caption{Error as a function of the cut-off heights $h_A, h_B$ for a pair of simulated regions for four simulation scenarios, with a true inter-correlation $\rho^{A,B}=0.3$. 
    The yellow diamond represents the error for cut-off heights equal to the maximum distance between U-scores within each region. The orange point corresponds to the minimal error.} 
    \label{fig:MSE_simu}
\end{figure}

The resulting surfaces are displayed in Figure \ref{fig:MSE_simu}. The lower the error, the better the quality of the estimator. As expected from Theorems \ref{th:ineq.h} and \ref{th:consistent}, the error is lowest (refer to the orange points in Figure \ref{fig:MSE_simu}) for cut-off heights that are neither too small nor too large. 
Moreover, when both the intra-correlation and the variance of the noise are low, the error is low, even for low cut-off heights, as there is no need to aggregate the data to obtain a consistent estimator.
However, the error is high for large cut-off heights regardless of the scenario. Indeed, even in the high noise settings, 
intra-correlation still influences the inter-correlation, and this effect is compounded by that of the cluster size.

In Section \ref{sec:clustering}, we proposed a computationally cheap heuristic to determine a suitable cut-off height.
Empirically, it seems the maximum distance between U-scores within a given region $A$, $h_A^{\text{max}}$, could indeed be an optimal cut-off height. It is represented by a yellow diamond in Figure \ref{fig:MSE_simu}. In fact, it seems to be located at the bottom of a valley and quite close to the minimal error for all settings.

\begin{figure}[b!]
        \begin{subfigure}[b]{.44\textwidth}
            \centering
            \includegraphics[width=\textwidth]{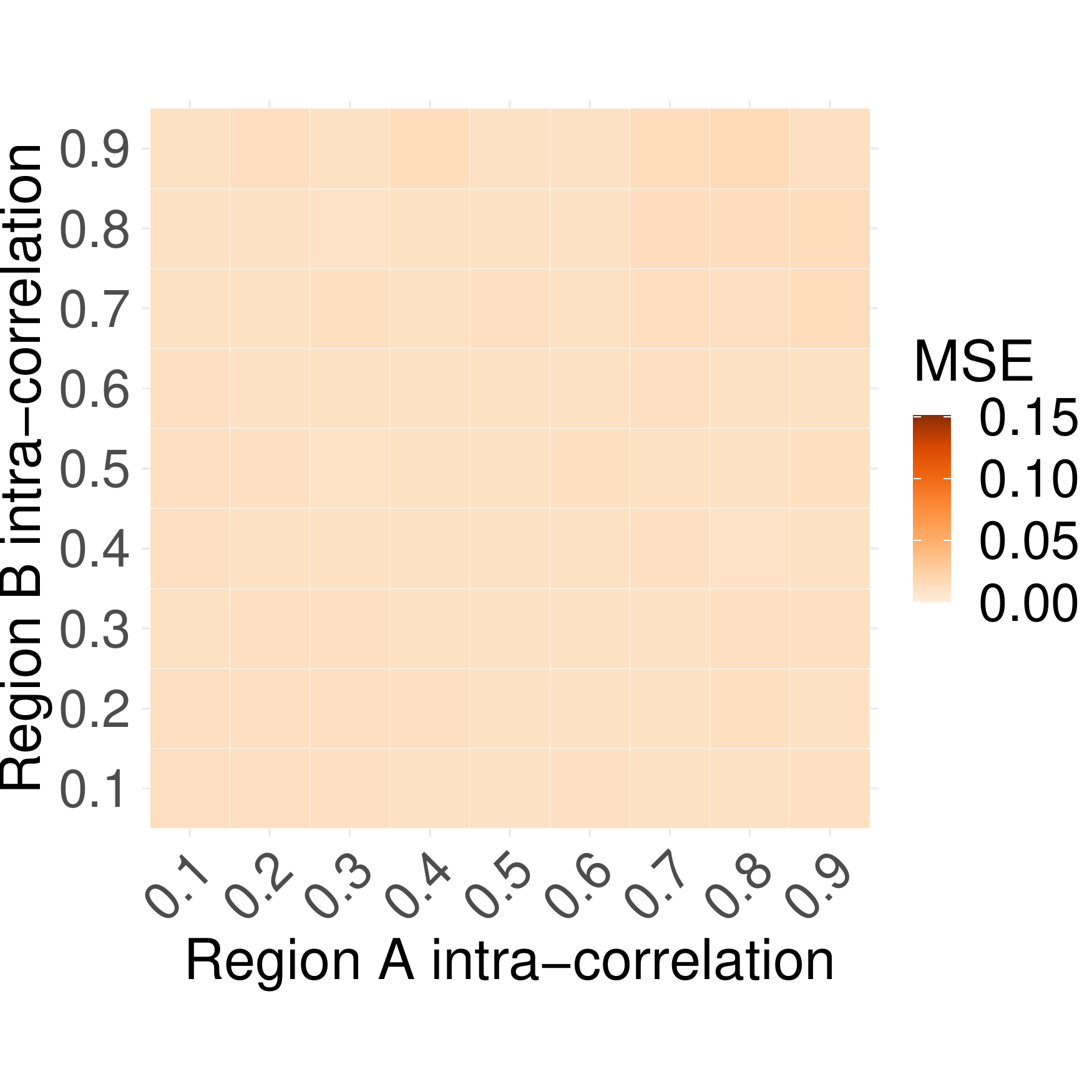}
            \caption{Maximum distance between U-scores.}
        \end{subfigure}
        ~
        \begin{subfigure}[b]{.44\textwidth}
            \centering
            \includegraphics[width=\textwidth]{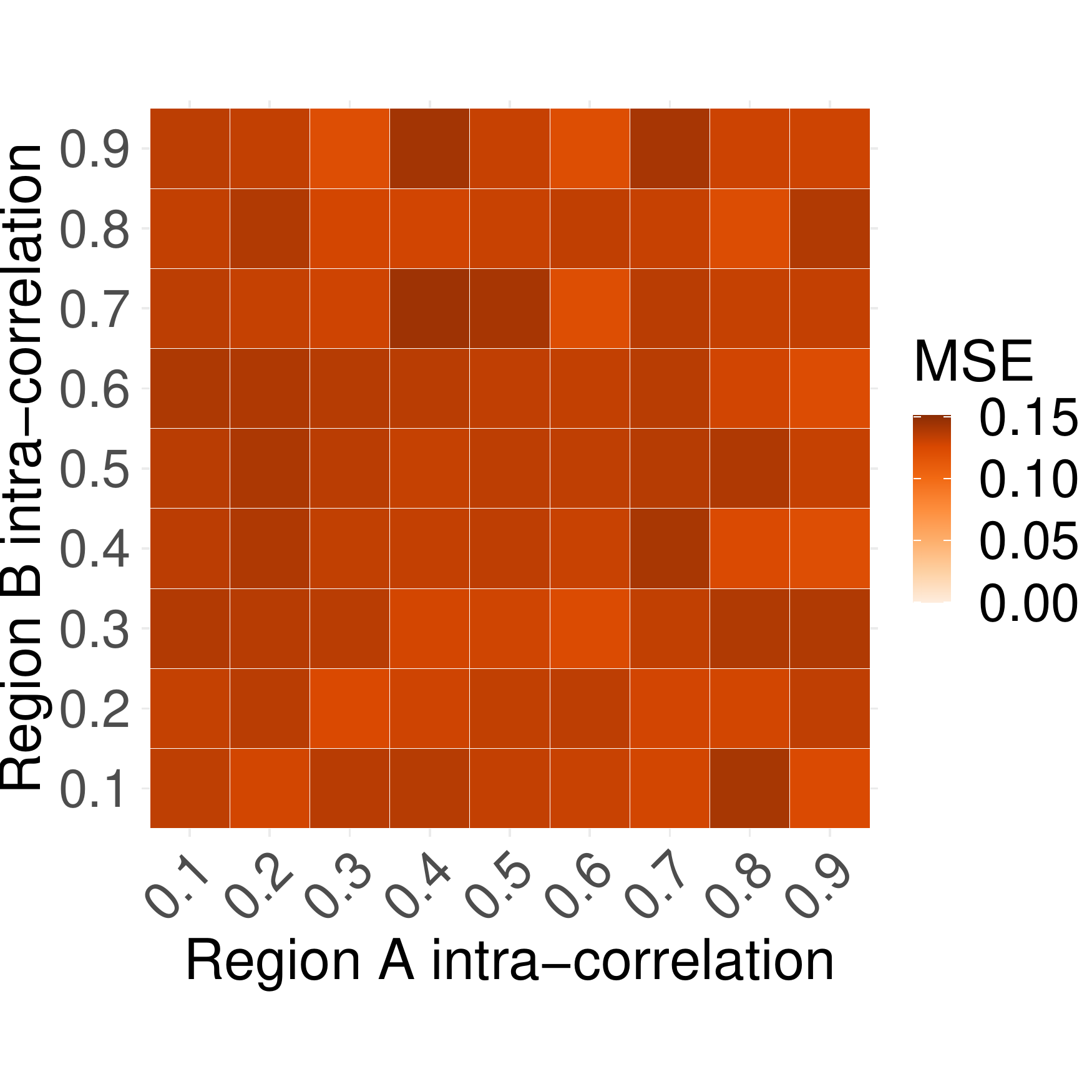}
            \caption{Maximum silhouette score}
        \end{subfigure}
        \caption{MSE ($\times 10$), averaged over 50 replicates, for varying intra-correlation values for regions $A$ and $B$. The true inter-correlation $\rho_{A,B}$ is $0.3$ and the noise variance $\sigma_{ {\displaystyle \epsilon}^{A}}^2= \sigma_{ {\displaystyle \epsilon}^{B}}^2 = 0.5$.}
        \label{fig:mse_criteria1}
    \end{figure}

\begin{figure}[t!]
        \begin{subfigure}[b]{.44\textwidth}
            \centering
            \includegraphics[width=\textwidth]{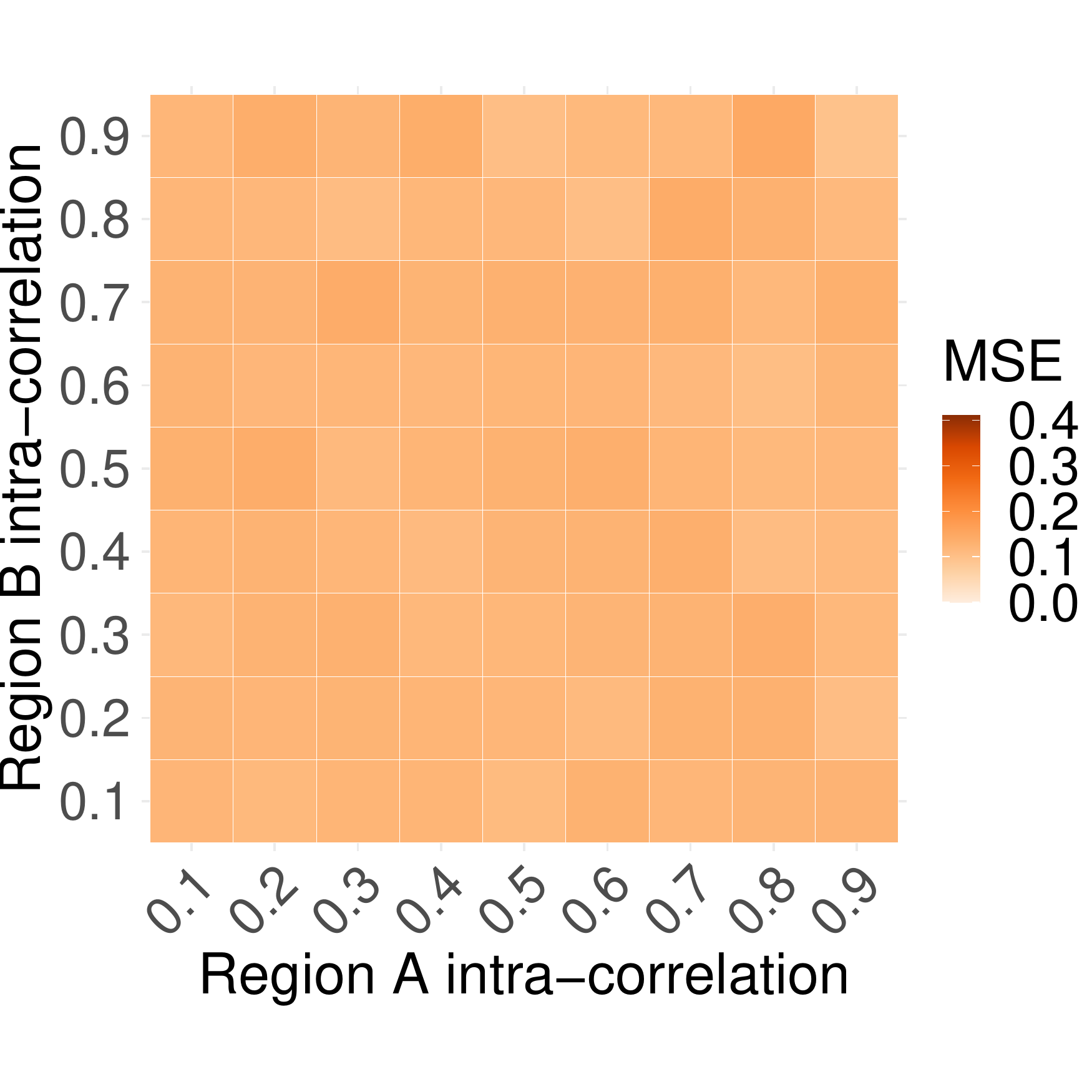}
            \caption{Maximum distance between U-scores.}
        \end{subfigure}
        ~
        \begin{subfigure}[b]{.44\textwidth}
            \centering
            \includegraphics[width =\textwidth]{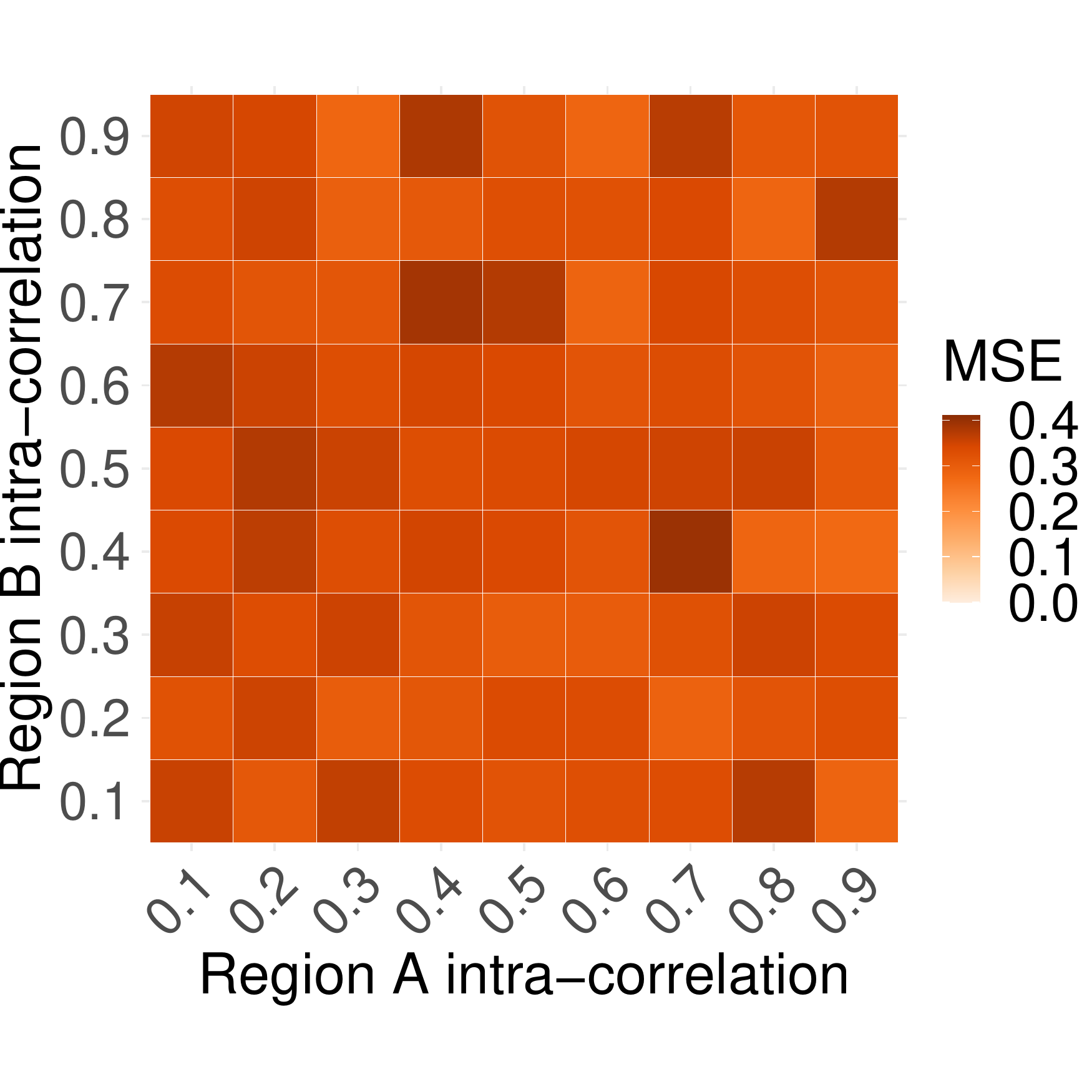}
            \caption{Maximum silhouette score.}
        \end{subfigure}
        \caption{MSE ($\times 100$), averaged over 50 replicates, for varying intra-correlation values for regions $A$ and $B$. The true inter-correlation $\rho_{A,B}$ is $0.1$ and the noise variance $\sigma_{ {\displaystyle \epsilon}^{A}}^2= \sigma_{ {\displaystyle \epsilon}^{B}}^2 = 0.5$.}
        \label{fig:mse_criteria2}
\end{figure}

We then compare our proposed optimal cut-off height, in terms of Mean Squared Error (MSE), to that obtained using a more standard criterion from the clustering literature: the maximum silhouette score. The Squared Error (SE) of a simulation-specific correlation estimate $r^{CLA}_{A,B}$ can be defined as
\begin{equation}
    \text{SE}=(r^{CLA}_{A,B} - \rho^{A,B} )^2.
\end{equation}
In this section, the MSE is computed by averaging the SEs across $50$ replicates. The MSE for varying intra- and inter-correlation values and a fixed high noise variance are depicted in Figures \ref{fig:mse_criteria1} and \ref{fig:mse_criteria2}. The MSE is lower when using our proposed cut-off heights in all the considered scenarios. 


From now on, and unless stated otherwise, we will hence estimate the inter-correlation using this optimal cut-off height.

\subsection{Comparison With Other Methods} 
\label{sec:choiceMethod}

We then empirically evaluate our choice of clustering method and compare our proposed approach with other estimators in terms of MSE. 

We first compare 
the performance of hierarchical clustering with Ward's linkage (our proposed choice and later denoted WardMaxU) with that of k-means \citep{kmeans} and ClustOfVar (CoV) \citep{chavent_clustofvar_2012}. ClustofVar is a hierarchical clustering method which is based on a principal component analysis approach, and closely related to works from \cite{dhillon_diametrical_2003} and \cite{vigneau_clustvarlv_2015}. DBSCAN \citep{dbscan1996}, which allows to directly control the cluster radii, was also considered. However, it fails to produce any clustering on the type of data we handle, 
which is high-dimensional. We also compare these clustering methods with a random assignment of the voxels into clusters (Random). We choose the cut-off heights required by Ward's method according to the heuristic validated in the previous section (that is the maximum distance between U-scores). ClustOfVar, k-means and Random all require a choice of the number of clusters (and not of the cut-off heights). We hence choose the former 
as that obtained with our proposed method. We also evaluate ClustOfVar with the number of clusters chosen according to the maximum rand index 
(randCoV), which is the proposed criterion in \citep{chavent_clustofvar_2012}. Results are presented in Table \ref{tab:MSE_simu_clustering}. All methods with the same number of clusters are similar, with the exception of the random assignment. As expected, the latter displays MSEs an order of magnitude higher than that of the other clustering techniques, except when both minimal intra-correlations are high. Indeed, in such cases, the intra-correlation is high enough that the intra-cluster correlation will be high regardless of the choice of clusters. This demonstrates the importance of constructing clusters with high intra-cluster correlation to correctly estimate the inter-correlation. 
The method randCoV showcases the second highest MSE in all scenarios, except when both intra-correlation and noise are high, in which case its MSE is similar to that of the k-means and CoV. 
Moreover, the computation of the rand index requires a bootstrapping step and is thus very 
computationally expensive. Indeed, the average CPU 
time of clustering two regions using the method randCov is in the order of $10$ min, while average CPU 
time is approximately $5$ s when using CoV, $300$ ms 
using kmeans, and $30$ ms using WardMaxU. Additionally, neither k-means nor CoV provide any obvious theoretical guarantees on the intra-correlation values within each cluster. Furthermore, they require to compute the U-scores, unlike our method. Indeed, our approach only depends on the distance between U-scores, which can be obtained directly from the sample 
voxel-to-voxel 
inter-correlation coefficients, without transforming the signals into U-scores. This step has a CPU 
time of about $15$ s 
per region. These methods are thus much more computationally heavy. This confirms the choice of hierarchical clustering with Ward's linkage for our purposes, and will be used in all subsequent results. 

\begin{table}[h!]
    \caption{Mean ($\times 10^{-3}$) and standard deviation in parenthesis ($\times 10^{-3}$) of the squared errors over 50 replicates for different clustering methods and different simulation scenarios from the 1D Toeplitz model. The inter-correlation $\rho^{A,B}$ is set to $0.3$.} 
    \centering
    \begin{tabular}{ccc c c c c c}
    \hline
    \multicolumn{3}{c}{Scenarios} & \multicolumn{5}{c}{Clustering Methods}  \\
    \cmidrule(r){1-3} \cmidrule(r){4-8}  $\eta_A^{-}$ & $\eta_B^{-}$ & $\gamma_A^2= \gamma_B^2$ & K-means & CoV & randCoV & Random & \textbf{WardMaxU}  \\
    \hline
    $0.2$ & $0.2$ & $0.5$ & $2.0$ ($1.4$) & $2.0$ ($1.4$) & $4.8$ ($7.8$) & $15$ ($5.2$) & $2.0$ ($1.4$) \\
    $0.8$ & $0.8$ & $0.5$ & $1.2$ ($1.5$) & $1.2$ ($1.5$) & $1.1$ ($1.3$) & $1.0$ ($1.0$) & $1.2$ ($1.5$) \\
     $0.2$ & $0.8$ & $0.5$ & $1.1 $ ($1.2$) & $1.1 $ ($1.2$) & $2.9$ ($4.2$) & $5.0$ ($3.1$) & $1.1$ ($1.2$) \\
    $0.2$ & $0.2$ & $0.1$ & $1.0$ ($0.9$) & $1.0$ ($0.9$) & $4.6$ ($10$) & $26$ ($8.1$) & $1.0$ ($0.9$) \\
    $0.8$ & $0.8$ & $0.1$ & $0.6$ ($1.0$) & $0.6$ ($1.1$) & $1.0$ ($1.4$) & $1.4$ ($1.6$) & $0.6$ ($1.1$) \\
    $0.2$ & $0.8$ & $0.1$ & $0.4$ ($0.6$) & $0.4$ ($0.5$) & $2.7$ ($4.4$) & $10$ ($4.5)$ & $0.4$ ($0.5$) \\
    \hline
    \end{tabular}
    \label{tab:MSE_simu_clustering}
\end{table}

We then compare our proposed estimator with the 
standard correlation of averages estimator $r^{CA}_{A,B}$, and the average of correlations $r^{AC}_{A,B}$ \citep{rosner.1977.1}. We also conduct comparisons with another inter-correlation estimator from the familial data literature, which is specifically designed for groups of dependent variables but fails to take into account noise \citep{Elston1975Cor}. Its quality is similar to that of $r^{AC}_{A,B}$, and these results are hence included in the supplementary materials. Comparison with other correlation estimators from the literature would not be fair as they either only consider 
pairs of variables or do not handle arbitrary inter-correlation.
To proceed 
we compute the regional-level point estimator $r^{CLA}_{A,B}$. We then calculate the MSE across $50$ simulations. The results obtained for several simulation scenarios are recorded in Table \ref{tab:MSE_simu}. As expected from Theorem \ref{th:consistent} and its 
corollary, our proposed estimator $r^{CLA}_{A,B}$ outperforms the other estimators for all settings, except the low noise scenarios with 3D Spherical intra-correlation, where the MSE for $r^{AC}_{A,B}$ is slightly lower. Even in this case, the MSE for $r^{AC}_{A,B}$ and $r^{CLA}_{A,B}$ are in the same order of magnitude. More generally, we can note that in all scenarios where the intra-correlation is quite high and 
the noise variance is low, the MSE for these two estimators are also in the same order of magnitude. Indeed, according to equation \eqref{eq:vox}, Theorem \ref{th:ineq.h}, and Corollary \ref{cor:consistency} 
$r^{AC}_{A,B}$ and $r^{CLA}_{A,B}$ would be very similar.
Therefore, not only is the quality of the estimation greatly improved in the presence of noise and low intra-correlation, but it is also not deteriorated when intra-correlation is high and the noise is low. Furthermore, in practice, data are expected to be quite noisy with a low intra-correlation.

We can remark here that we did not include in Table \ref{tab:MSE_simu} scenarios where the intra-correlation is close to zero. Indeed, in such cases no clusters of highly correlated variables can be found. In practical situations, this could be due to either high regional inhomogeneity or high noise, and could indicate an issue with the parcellation or data acquisition. Our clustering approach can hence help identify problematic datasets and thus provide information on the quality of the data.

\begin{table}[h!]
    \caption{Mean and standard deviation (in parenthesis) of the squared error over $50$ replicates for different simulation scenarios 
    and different estimators. The inter-correlation $\rho^{A,B}$ is set to $0.3$. 
    }
    \centering
    \resizebox{\textwidth}{!}{
    \begin{tabular}{c|ccc|c|c|c}
    \hline
    \multicolumn{4}{c}{Scenarios} & \multicolumn{3}{c}{Estimators}  \\
    \cmidrule(r){1-4} \cmidrule(r){5-7} 
    \multirow{7}{*}{\rotatebox[origin=r]{90}{1D Toeplitz}} & $\eta_A^{-}$ & $\eta_B^{-}$ & $\gamma_A^2, \gamma_B^2$ &  $r^{AC}_{A,B}$ & $r^{CLA}_{A,B}$ & $r^{CA}_{A,B}$\\
    \cmidrule(r){2-7}
      &  $0.2$ & $0.2$ & $0.5$ & $1.8\times 10^{-2}$ ($2.8\times10^{-3}$) & $\mathbf{2.0\times 10^{-3}}$ ($1.4\times10^{-3}$) & $1.5\times 10^{-1}$ ($1.8\times10^{-1}$) \\
    & $0.8$ & $0.8$ & $0.5$ & $1.2\times 10^{-2}$ ($3.7\times10^{-3}$) & $\mathbf{1.2 \times 10^{-3}}$ ($1.5\times10^{-3}$) & $1.0\times 10^{-1}$ ($1.0\times10^{-1}$) \\
    & $0.2$ & $0.8$ & $0.5$ & $1.4\times 10^{-2}$ ($3.0\times10^{-3}$) & $\mathbf{1.1 \times 10^{-3}}$ ($1.2\times10^{-3}$) & $1.0\times 10^{-1}$ ($1.0\times10^{-1}$)\\
    & $0.2$ & $0.2$ & $0.1$ & $5.4\times 10^{-3}$ ($2.0\times10^{-3}$) & $\mathbf{1.0\times 10^{-3}}$ ($9.1\times10^{-4}$) & $2.3\times 10^{-1}$ ($2.7\times10^{-1}$) \\
    & $0.8$ & $0.8$ & $0.1$ & $1.9\times 10^{-3}$ ($2.0\times10^{-3}$) & $\mathbf{6.4\times 10^{-4}}$ ($1.0\times10^{-3}$) & $1.2\times 10^{-1}$ ($1.2\times10^{-1}$) \\
    & $0.2$ & $0.8$ & $0.1$ & $2.7\times 10^{-3}$ ($1.7\times10^{-3}$) & $\mathbf{4.3\times 10^{-4}}$ ($5.5\times10^{-4}$) & $1.4\times 10^{-1}$ ($1.6\times10^{-1}$) \\
    \hline
    \hline
    \multirow{7}{*}{\rotatebox[origin=r]{90}{3D Matérn}} & $\phi_{A,A}$ & $\phi_{B,B}$ & $\gamma_A^2, \gamma_B^2$ &  $r^{AC}_{A,B}$ & $r^{CLA}_{A,B}$ & $r^{CA}_{A,B}$\\
    \cmidrule(r){2-7}
    & $0.6$ & $0.6$ & $0.5$ & $1.0 \times 10^{-2} $ ($3.8 \times 10^{-3}$) & $\mathbf{7.0 \times 10^{-4}}$ ($1.1 \times 10^{-3}$)  & $1.6 \times 10^{-3}$ ($1.9 \times 10^{-3}$) \\
    & $0.8$ & $0.8$ & $0.5$ & $1.0 \times 10^{-2}$ ($4.0 \times 10^{-3}$) & $\mathbf{7.9 \times 10^{-4}}$ ($1.2 \times 10^{-3}$) & $1.0 \times 10^{-3}$ ($1.4 \times 10^{-3}$) \\
    & $0.6$ & $0.8$ & $0.5$ & $1.0 \times 10^{-2}$ ($3.9 \times 10^{-3}$) & $\mathbf{7.2 \times 10^{-4}}$ ($1.1 \times 10^{-3}$) & $1.0 \times 10^{-3}$ ($1.6 \times 10^{-3}$) \\
    & $0.6$ & $0.6$ & $0.1$ &  $1.3 \times 10^{-3}$ ($1.5 \times 10^{-3}$) &  $\mathbf{7.7 \times 10^{-4}}$ ($1.0 \times 10^{-3}$) &  $1.7 \times 10^{-3}$ ($2.0 \times 10^{-3}$) \\
    & $0.8$ & $0.8$ & $0.1$ &  $1.4 \times 10^{-3}$ ($1.6 \times 10^{-3}$) &  $\mathbf{7.5 \times 10^{-4}}$ ($1.0 \times 10^{-3}$) &  $1.1 \times 10^{-3}$ ($1.4 \times 10^{-3}$) \\
    & $0.6$ & $0.8$ & $0.1$ &  $1.3 \times 10^{-3}$ ($1.6 \times 10^{-3}$) &  $\mathbf{7.7 \times 10^{-4}}$ ($1.0 \times 10^{-3}$) &  $1.3 \times 10^{-3}$ ($1.7 \times 10^{-3}$) \\
    \hline
    \hline 
    \multirow{7}{*}{\rotatebox[origin=r]{90}{3D Spherical}} & $\phi_{A,A}$ & $\phi_{B,B}$ & $\gamma_A^2, \gamma_B^2$ &  $r^{AC}_{A,B}$ & $r^{CLA}_{A,B}$ & $r^{CA}_{A,B}$\\
    \cmidrule(r){2-7}
    & $8$ & $8$ & $0.5$ & $1.0 \times 10^{-2} $ ($2.3 \times 10^{-3}$) & $\mathbf{4.6 \times 10^{-3}}$ ($2.4 \times 10^{-3}$)  & $8.8 \times 10^{-2}$ ($1.4 \times 10^{-2}$) \\
    & $12$ & $12$ & $0.5$ & $1.0 \times 10^{-2}$ ($2.8 \times 10^{-3}$) & $\mathbf{2.4 \times 10^{-3}}$ ($1.9 \times 10^{-3}$) & $2.5 \times 10^{-2}$ ($8.2 \times 10^{-3}$) \\
    & $8$ & $12$ & $0.5$ & $9.4 \times 10^{-3}$ ($2.5 \times 10^{-3}$) & $\mathbf{4.2 \times 10^{-3}}$ ($2.3 \times 10^{-3}$) & $5.3 \times 10^{-2}$ ($1.1 \times 10^{-2}$) \\
    & $8$ & $8$ & $0.1$ &  $\mathbf{9.1 \times 10^{-4}}$ ($7.9 \times 10^{-4}$)  &  $8.9 \times 10^{-3}$ ($3.8 \times 10^{-3}$) &  $9.3 \times 10^{-2}$ ($1.3 \times 10^{-2}$) \\
    & $12$ & $12$ & $0.1$ & $\mathbf{1.0 \times 10^{-3}}$ ($1.0 \times 10^{-3}$) &  $4.5 \times 10^{-3}$ ($2.8 \times 10^{-3}$) &  $2.6 \times 10^{-2}$ ($8.4 \times 10^{-3}$) \\
    & $8$ & $12$ & $0.1$ &  $\mathbf{7.3 \times 10^{-4}}$ ($7.8 \times 10^{-4}$) &  $7.7 \times 10^{-3}$ ($3.3 \times 10^{-3}$) &  $5.6 \times 10^{-2}$ ($1.1 \times 10^{-2}$) \\
    \hline
    \end{tabular}
    } 
    \label{tab:MSE_simu}
\end{table}

\subsection{Illustration on Real-world Data}

We now apply 
our proposed estimator on real-world fMRI datasets, with the goal of estimating functional connectivity. At first, the sample 
cluster-level inter-correlation and voxel-level intra-correlation of different subjects can be visually inspected. The correlation estimates of three rats, including a dead one, are displayed in Figure \ref{fig:corrmat_rat}, and that of three healthy human subjects (from the HCP dataset) are shown in Figure \ref{fig:corrmat_HCP}.

In brain functional connectivity studies, point estimates for each pair of regions are needed to construct a correlation matrix. A thresholding step is then 
applied to obtain a binary connectivity network where only the edges corresponding to the highest correlation values remain. In this section, we will therefore 
mostly focus on evaluating the 
regional-level entries of these correlation matrices. 

\subsubsection{Rat Data}

\begin{figure}[th!]
    \centering
    \begin{subfigure}{0.3\textwidth}
			\centering
			\includegraphics[width=.9\linewidth]{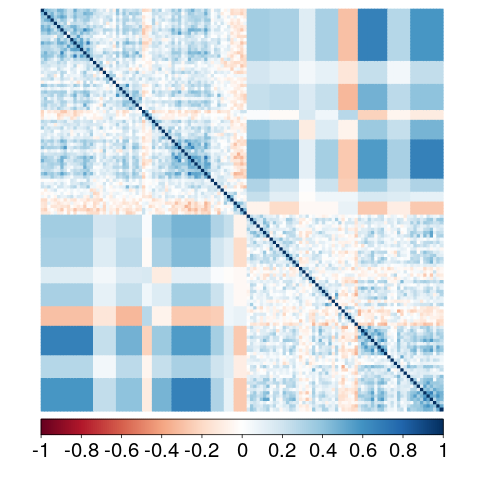}
			\caption{M1\_l, M1\_r-rat 24 (IsoW)}
		\end{subfigure}
		~
		\begin{subfigure}{0.3\textwidth}
			\centering
			\includegraphics[width=.9\linewidth]{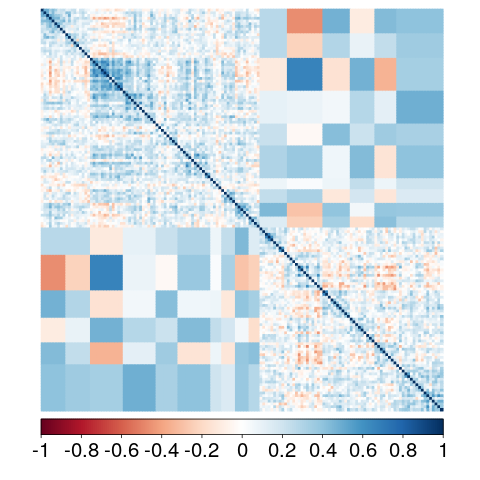}
			\caption{M1\_l, M1\_r--rat 31 (EtoL)}
		\end{subfigure}
		~
		\begin{subfigure}{0.3\textwidth}
			\centering
			\includegraphics[width=.9\linewidth]{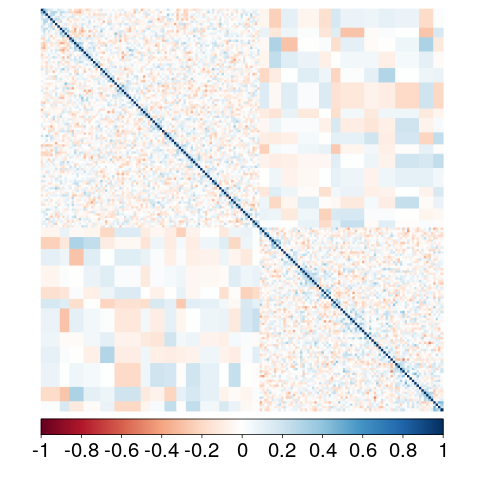}
			\caption{M1\_l, M1\_r--rat 9 (dead)}
		\end{subfigure}
        ~
		\begin{subfigure}{0.3\textwidth}
			\centering
			\includegraphics[width=.9\linewidth]{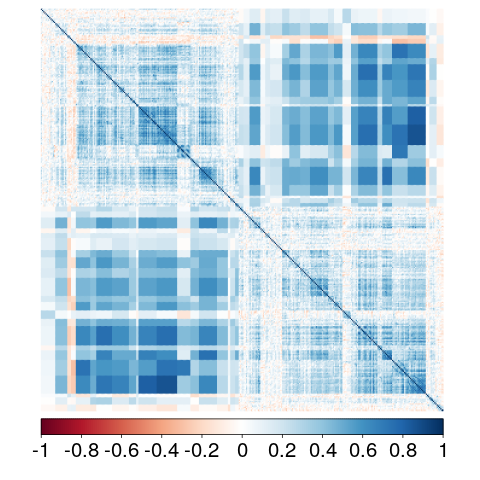}
			\caption{S1\_l, S1\_r--rat 24 (IsoW)}
		\end{subfigure}
        ~
		\begin{subfigure}{0.3\textwidth}
			\centering
			\includegraphics[width=.9\linewidth]{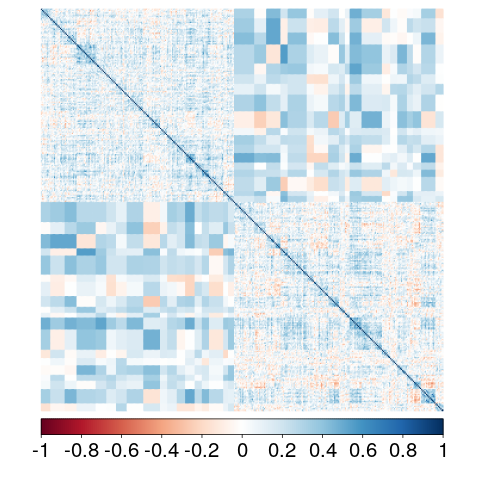}
			\caption{S1\_l, S1\_r--rat 31 (EtoL)}
		\end{subfigure}
		~
		\begin{subfigure}{0.3\textwidth}
			\centering
			\includegraphics[width=.9\linewidth]{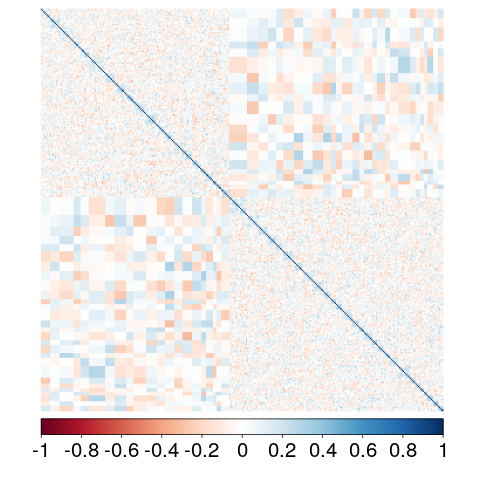}
			\caption{S1\_l, S1\_r--rat 9 (dead)}
		\end{subfigure}
    \caption{Sample pairwise correlation matrices for different rats and brain region pairs. Voxels are ordered by clusters. The diagonal blocks correspond to the voxel-to-voxel sample 
    intra-correlation $r^{A,A}_{i,i^{\prime}}$, while the off-diagonal blocks correspond to the sample 
    inter-correlation between clusters $r^{CLA}_{\nu_A,\nu_B}$. }
    \label{fig:corrmat_rat}
\end{figure}

\paragraph{Dead Rats}  No functional activity should be detected in dead rats, unlike in live 
rats. 
Dead rats hence provide experimental data where the ground-truth inter-correlation is zero. We can therefore compute the MSE 
across all pairs of regions (each region pair is a replicate). We expect 
as well that the intra-correlation is zero within all regions. In fact, no discernible structure of the dead rat's intra-correlation can 
be noted in Figure \ref{fig:corrmat_rat}, where motor (M1\_l, M1\_r) and sensory (S1\_l, S1\_r) regions are represented. We find the MSE of $r^{CLA}_{A,B}$ is slightly higher than that of $r^{AC}_{A,B}$ (cf. Table \ref{tab:mse_rat}). 
Nonetheless, they are both very low and several orders of magnitude lower than the MSE of $r^{CA}_{A,B}$. 
This indicates that for dead rat data, $r^{CLA}_{A,B}$ displays similar quality to $r^{AC}_{A,B}$, and a considerable improvement over the standard $r^{CA}_{A,B}$.

\begin{table}[htbp]
    \caption{MSE across all pairs of regions for different dead rats and different estimators.}
    \centering
    \begin{tabular}{c c c c}
    \hline
    Dead Rat ID & $r^{AC}_{A,B}$ & $r^{CLA}_{A,B}$ & $r^{CA}_{A,B}$  \\
     \hline 
    16 & $5.2 \times 10^{-6}$ & $5.6 \times 10^{-5}$ & $1.3\times 10^{-2}$ \\
    18 & $4.7 \times 10^{-6}$ & $5.4 \times 10^{-5}$ & $1.3\times 10^{-2}$ \\
   9 & $5.7\times 10^{-6}$ & $6.0\times 10^{-5}$ & $1.3\times 10^{-2}$ \\
   \hline
    \end{tabular}
    \label{tab:mse_rat}
\end{table}

\begin{figure}[h!]
    \centering
    \begin{subfigure}{0.31\textwidth}
			\centering
			\includegraphics[width=\linewidth]{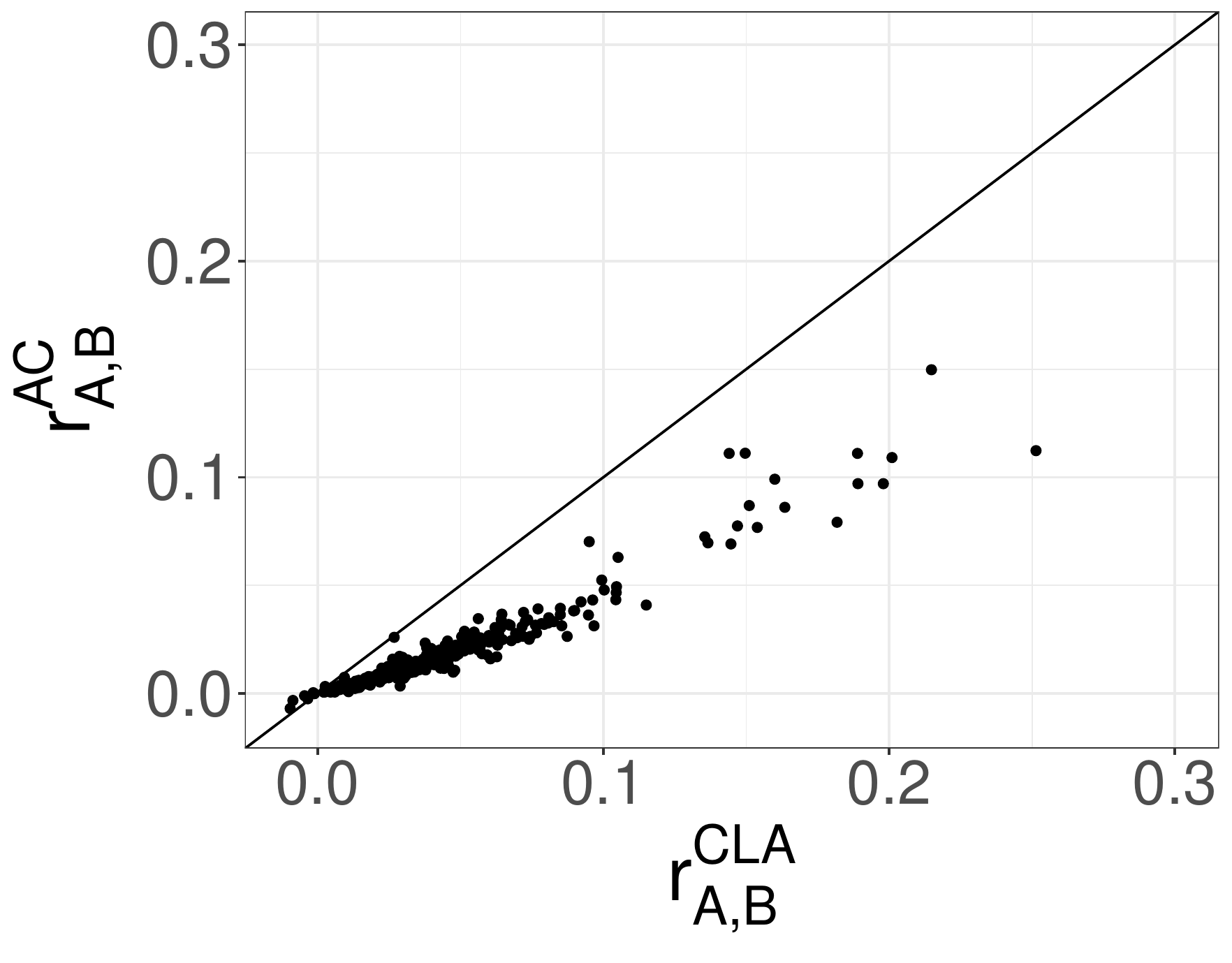} 
			\caption{Rat 24 (IsoW)}
	\end{subfigure}
	~
    \begin{subfigure}{0.31\textwidth}
			\centering
			\includegraphics[width=\linewidth]{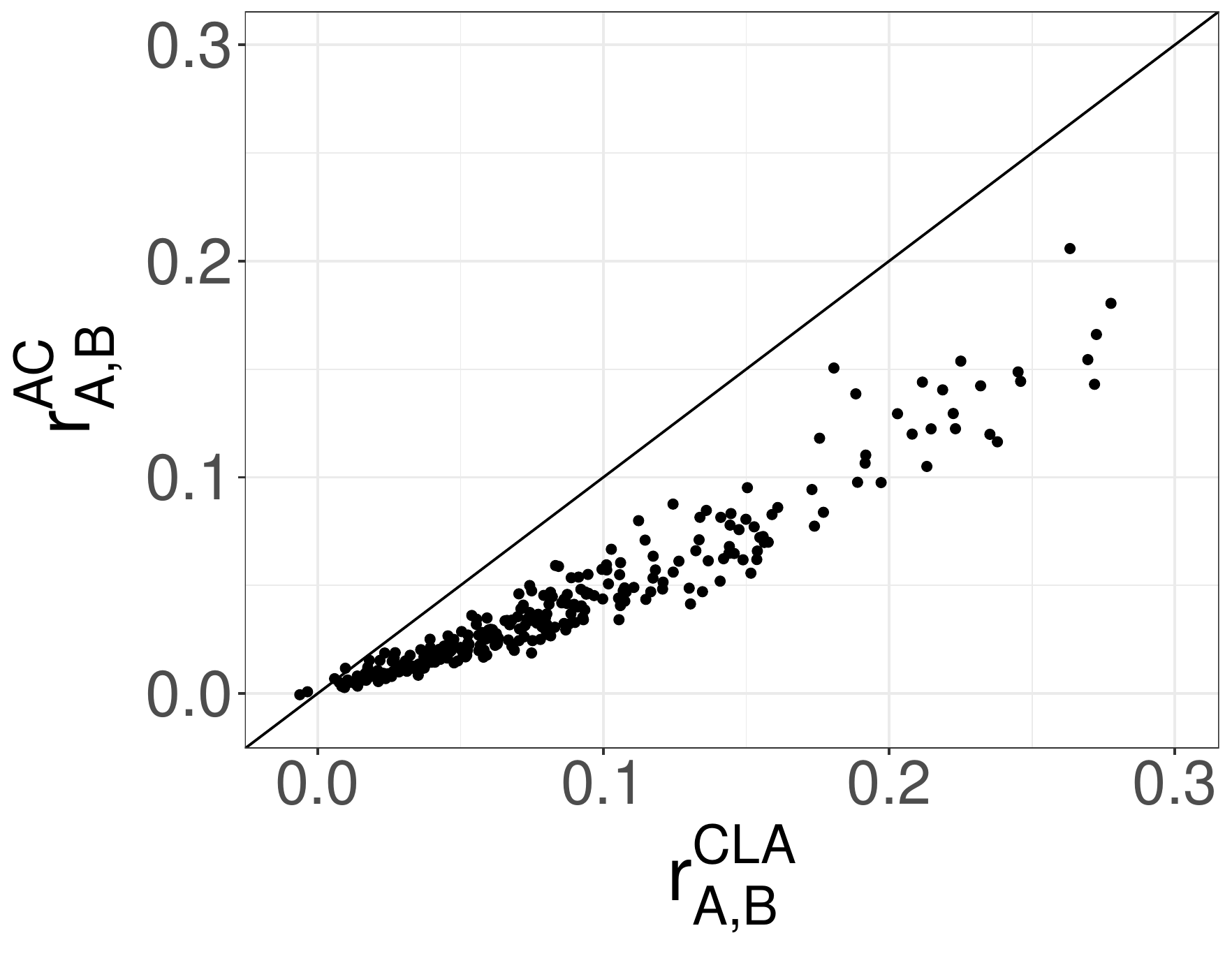} 
			\caption{Rat 
			4 (UreL)}
	\end{subfigure}
	~
	\begin{subfigure}{0.31\textwidth}
			\centering
			\includegraphics[width=\linewidth]{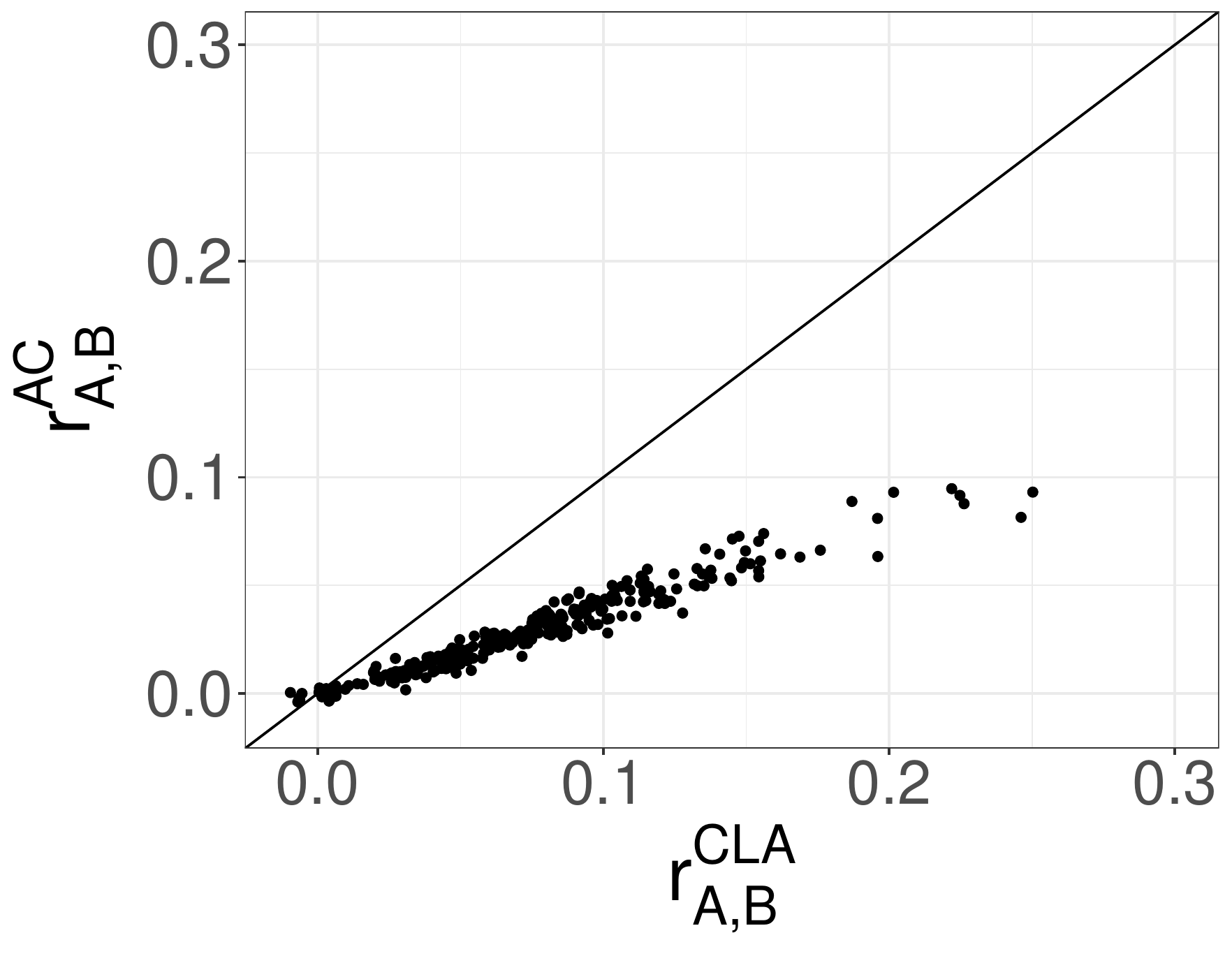} 
			\caption{Rat 31 
			(EtoL)}
	\end{subfigure}
	\caption{Sample inter-correlation coefficients estimated using $r^{AC}_{A,B}$ against our proposed estimator $r^{CLA}_{A,B}$ for three live rats under different anesthetics. Each point represents a pair of brain regions.} 
    \label{fig:scatter}
\end{figure}

\paragraph{Live Rats} To further illustrate the advantages of our proposed approach, we consider three live rats under different anesthetics. Unlike for dead rats, no ground-truth inter-correlation is available. We thus 
inspect directly the values of the estimated inter-correlations.
We can first remark correlation values are visually very different in live and dead rats. Indeed, both intra- and inter-correlations 
are higher, in addition to displaying an apparent structure (cf. Figure \ref{fig:corrmat_rat} ). 
While we could not clearly demarcate $r^{AC}_{A,B}$ from $r^{CLA}_{A,B}$ using solely the dead rat data, we can note in Figure \ref{fig:scatter} that for any pair of regions, $r^{CLA}_{A,B}$ is both larger than $r^{AC}_{A,B}$ and further away from zero, which corresponds to dead rat connectivity. In the context of functional connectivity, this implies that, when applying a thresholding step, 
$r^{CLA}_{A,B}$ may allow us to increase the number of rightfully detected edges in the corresponding connectivity network. 

\subsubsection{HCP Data} 
We then 
illustrate our proposed approach on human data from healthy live subjects. 
No ground-truth is available.

\begin{figure}[h!]
    \centering
    \begin{subfigure}{0.3\textwidth}
			\centering
			\includegraphics[width=.9\linewidth]{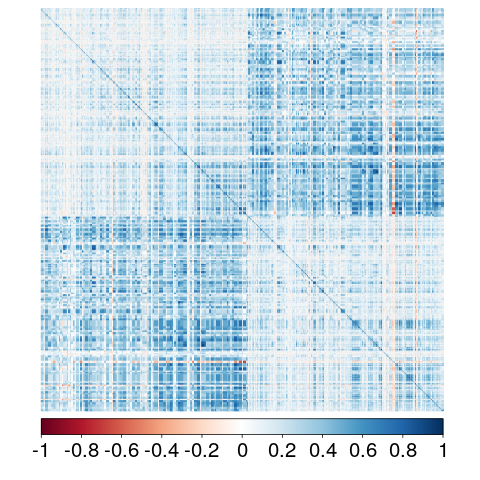}
			\caption{Pr\_l, Pr\_r -- subject 1}
		\end{subfigure}
		~
    \begin{subfigure}{0.3\textwidth}
			\centering
			\includegraphics[width=.9\linewidth]{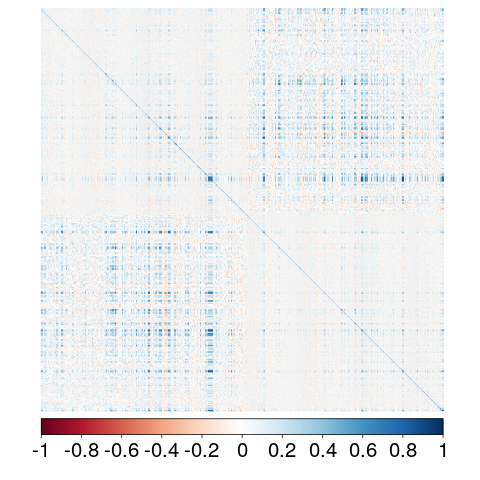}
			\caption{Pr\_l, Pr\_r -- subject 2}
		\end{subfigure}
		~
    \begin{subfigure}{0.3\textwidth}
			\centering
			\includegraphics[width=.9\linewidth]{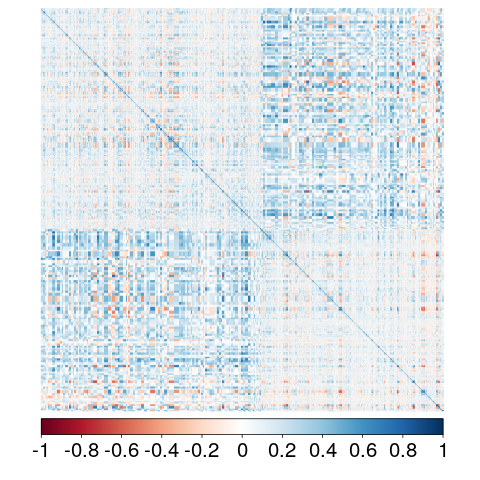}
			\caption{Pr\_l, Pr\_r -- subject 3}
		\end{subfigure}
        ~
		\begin{subfigure}{0.3\textwidth}
			\centering
			\includegraphics[width=.9\linewidth]{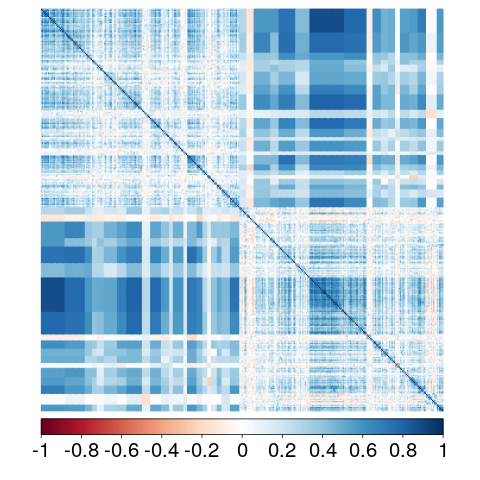}
			\caption{H\_l, H\_r -- subject 1}
		\end{subfigure}
        ~
		\begin{subfigure}{0.3\textwidth}
			\centering
			\includegraphics[width=.9\linewidth]{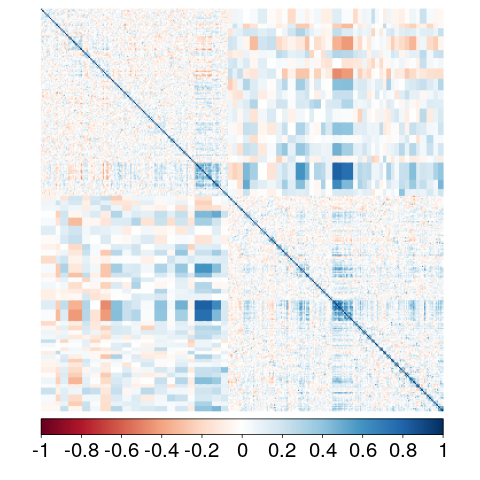}
			\caption{H\_l, H\_r -- subject 2}
		\end{subfigure}
		~
		\begin{subfigure}{0.3\textwidth}
			\centering
			\includegraphics[width=.9\linewidth]{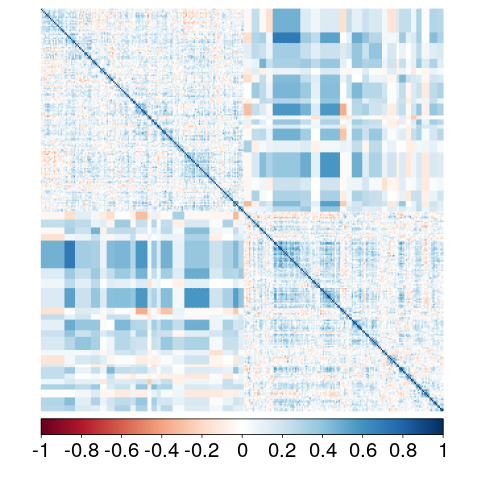}
			\caption{H\_l, H\_r -- subject 3 }
		\end{subfigure}
    \caption{Sample pairwise correlation matrices for different HCP subjects and brain region pairs. Voxels are ordered by clusters. The diagonal blocks correspond to the voxel-to-voxel sample 
    intra-correlation $r^{A,A}_{i,i^{\prime}}$, while the off-diagonal blocks correspond to the sample 
    inter-correlation between clusters $r^{CLA}_{\nu_A,\nu_B}$. }
    \label{fig:corrmat_HCP}
\end{figure}

Figure \ref{fig:corrmat_HCP} showcases 
sample correlations of the Precentral regions (Pr\_l, Pr\_r), which are large regions containing about $1700$ voxels, and Heschl's gyri (H\_l, H\_r), which are ten times smaller. We can first note 
that the intra-correlation displays some structure, as in the live rats. Nonetheless, overall, subject 2 seems to have both lower sample 
intra- and inter-correlation values, compared to most other subjects (including subjects 1 and 3). Subject 2 has in fact a benign anatomical brain anomaly. Our proposed approach hence allowed us to 
single out an 
unusual subject just by visually inspecting its sample intra- and inter-correlation values. 

\begin{figure}[h!]
    \centering
    \begin{subfigure}{0.31\textwidth}
			\centering
			\includegraphics[trim=0cm 0cm 0cm .7cm, clip=true, width=\linewidth]{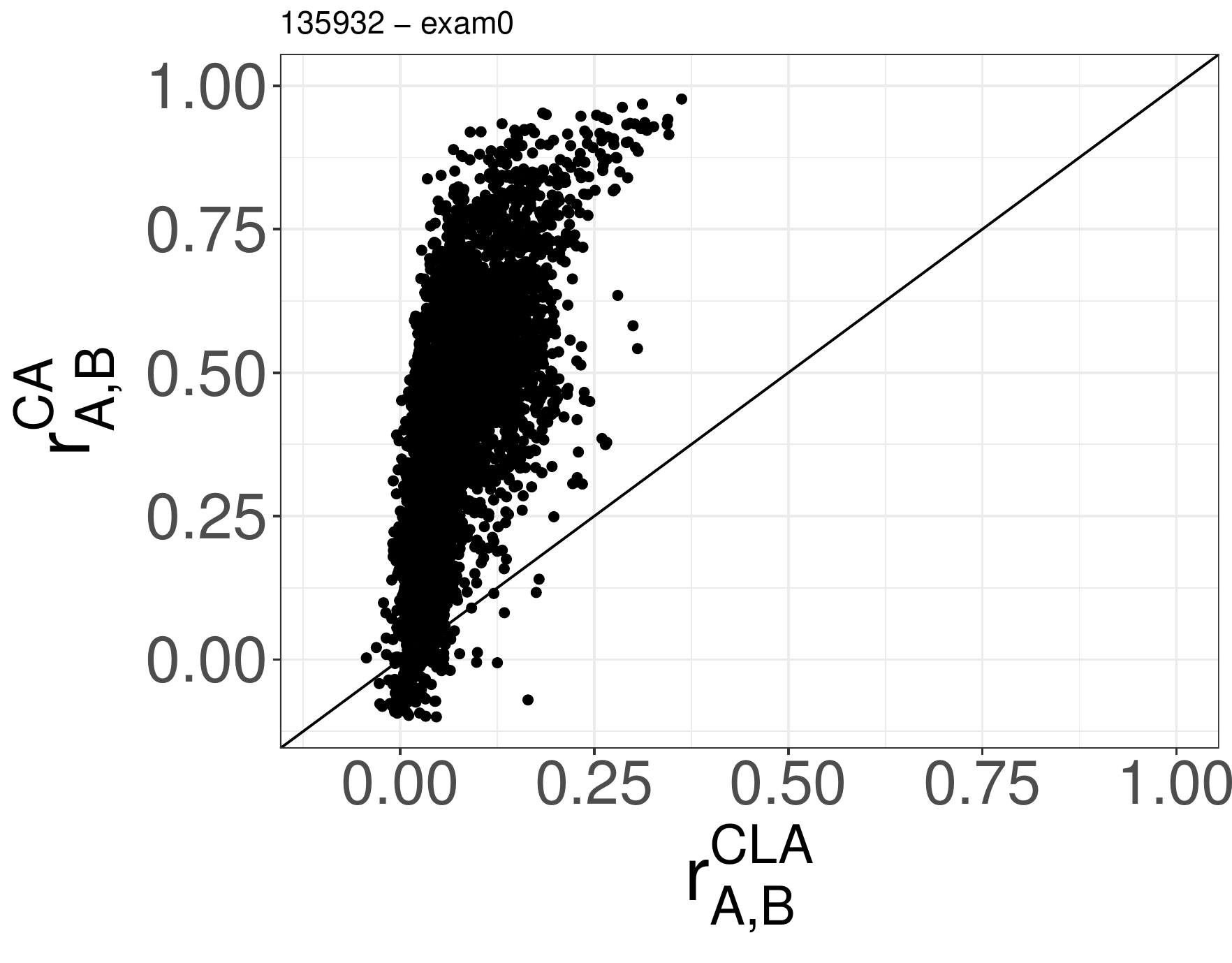} 
			\caption{Subject 1 -- session 0}
	\end{subfigure}
	~
    \begin{subfigure}{0.31\textwidth}
			\centering
			\includegraphics[trim=0cm 0cm 0cm .7cm, clip=true, width=\linewidth]{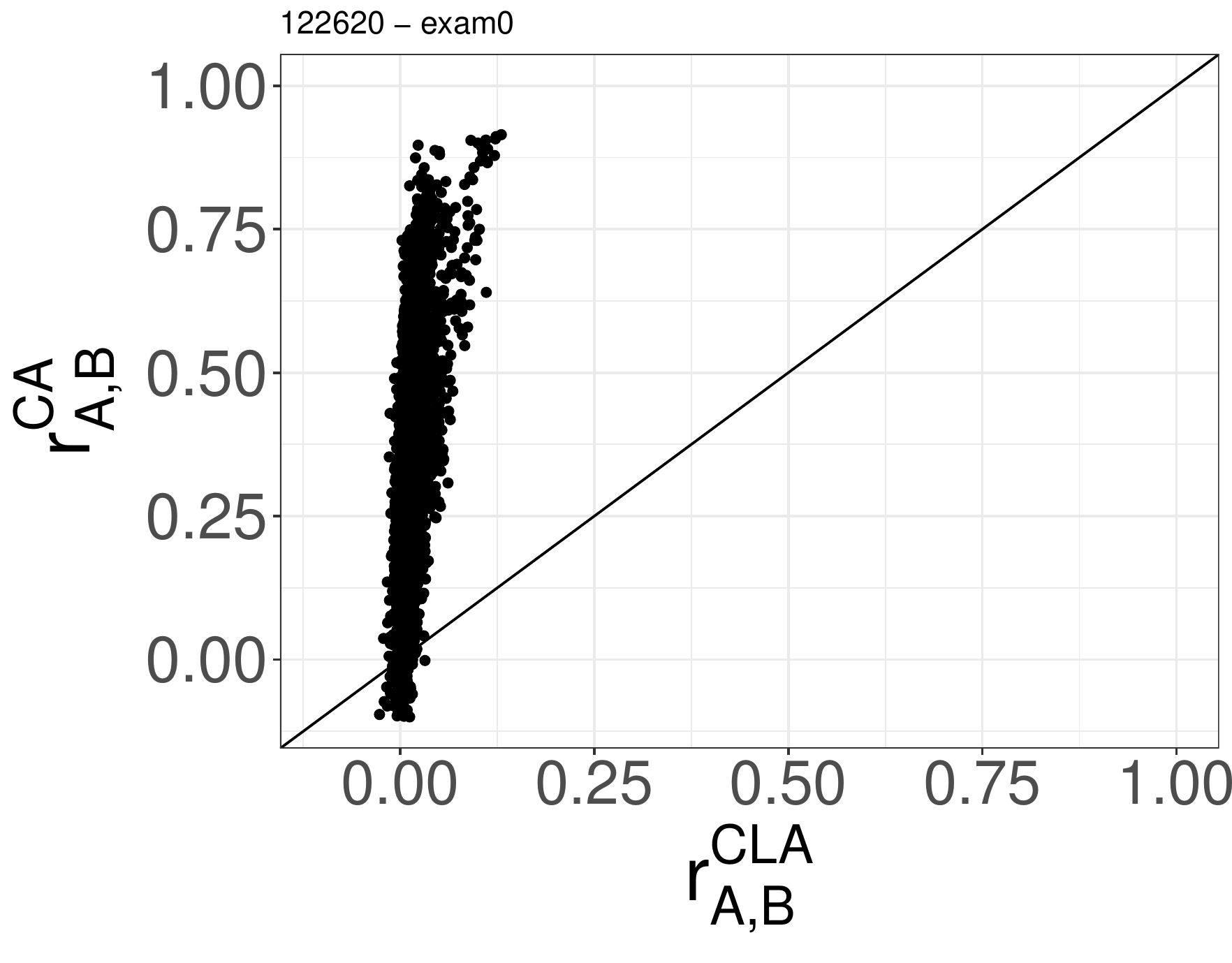} 
			\caption{Subject 2 -- session 0}
	\end{subfigure}
	~
    \begin{subfigure}{0.31\textwidth}
			\centering
			\includegraphics[trim=0cm 0cm 0cm .7cm, clip=true, width=\linewidth]{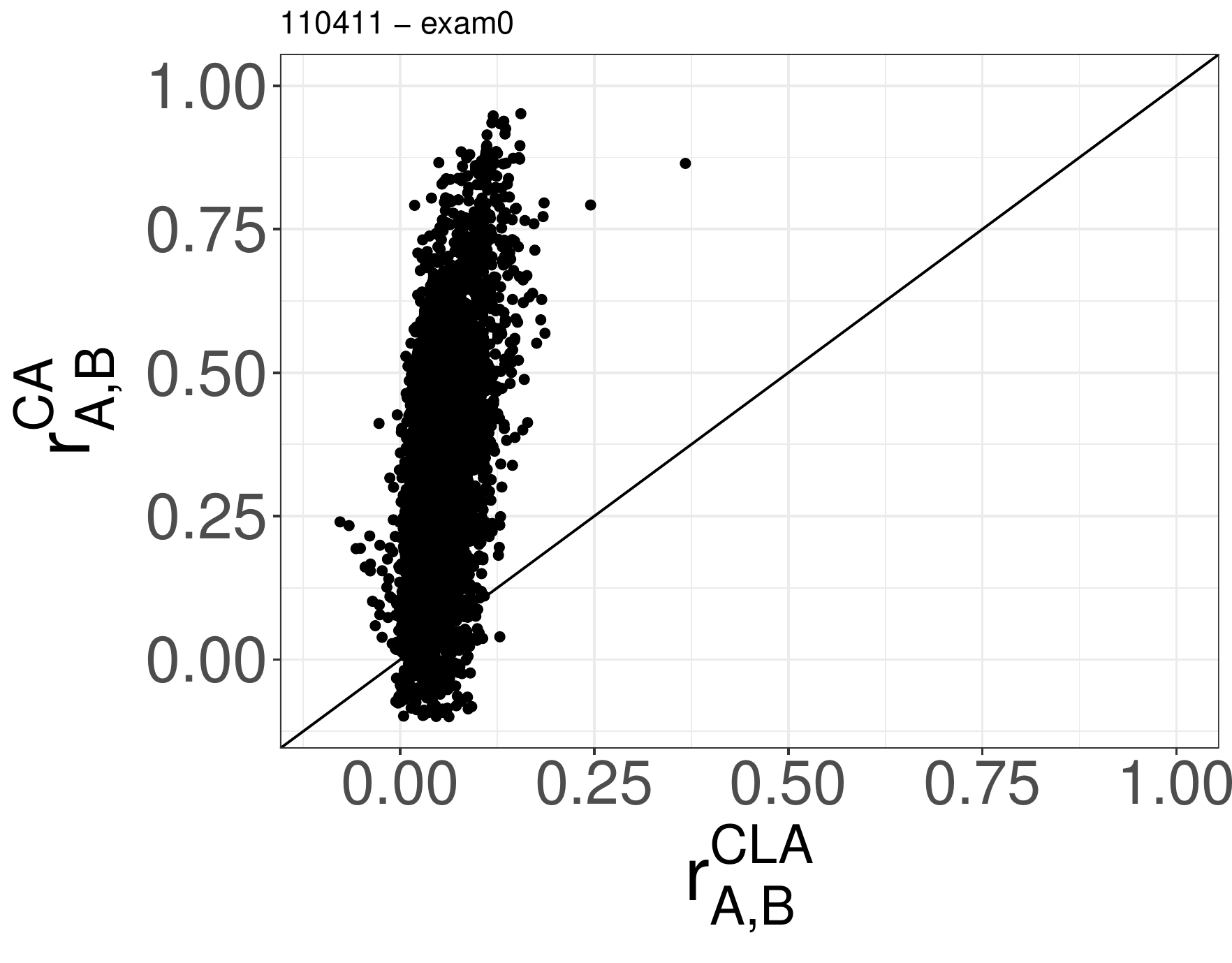} 
			\caption{Subject 3 -- session 0}
	\end{subfigure}
	\caption{
 Inter-correlation coefficients estimated using $r^{CA}_{A,B}$ against our proposed estimator $r^{CLA}_{A,B}$ for three HCP subjects. Each point represents a pair of brain regions.} 
    \label{fig:scatter_HCP_CA}
\end{figure}

\begin{figure}[h!]
    \centering
    \begin{subfigure}{0.31\textwidth}
			\centering
			\includegraphics[trim=0cm 0cm 0cm .7cm, clip=true, width=\linewidth]{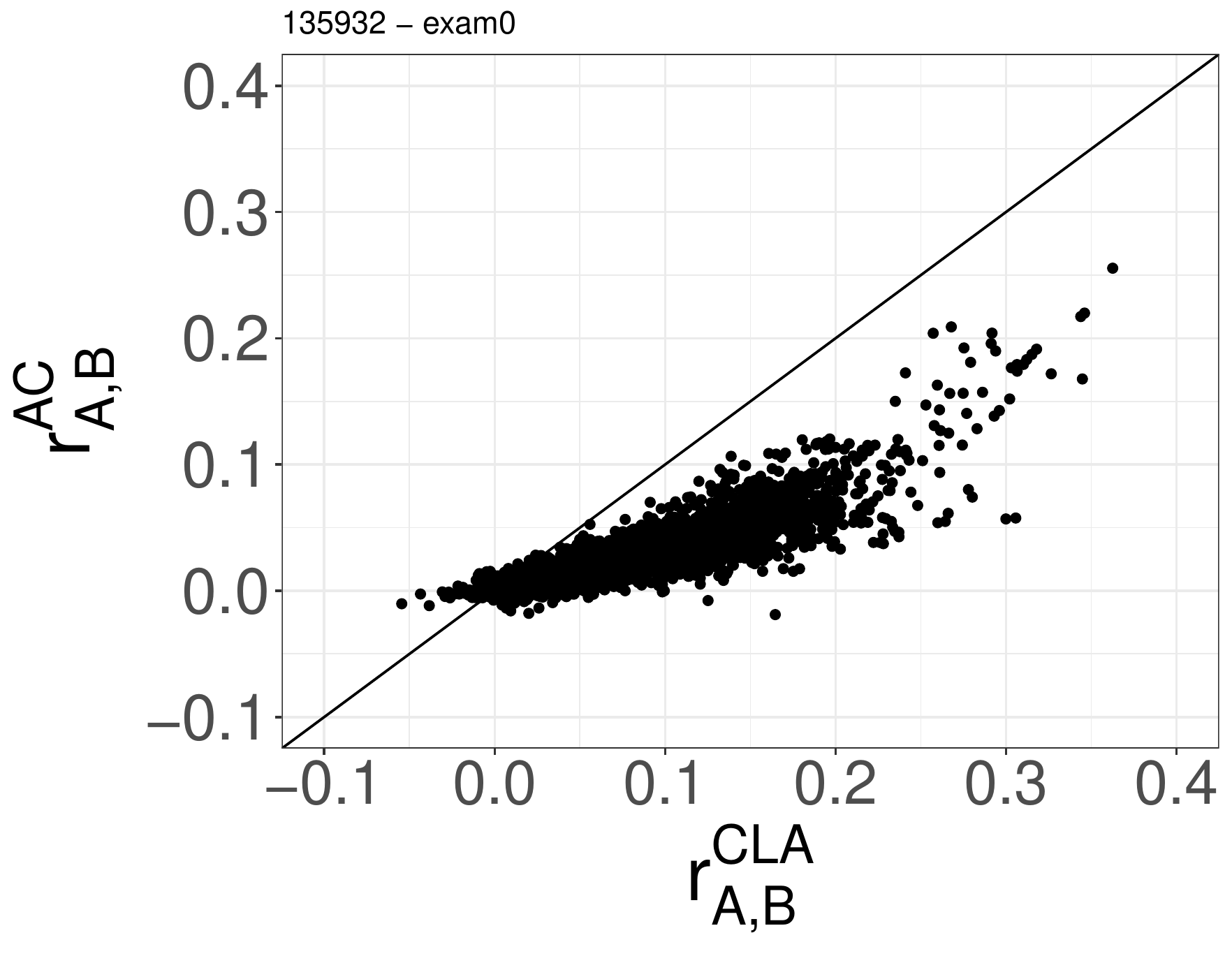} 
			\caption{Subject 1 -- session 0}
	\end{subfigure}
	~
    \begin{subfigure}{0.31\textwidth}
			\centering
			\includegraphics[trim=0cm 0cm 0cm .7cm, clip=true, width=\linewidth]{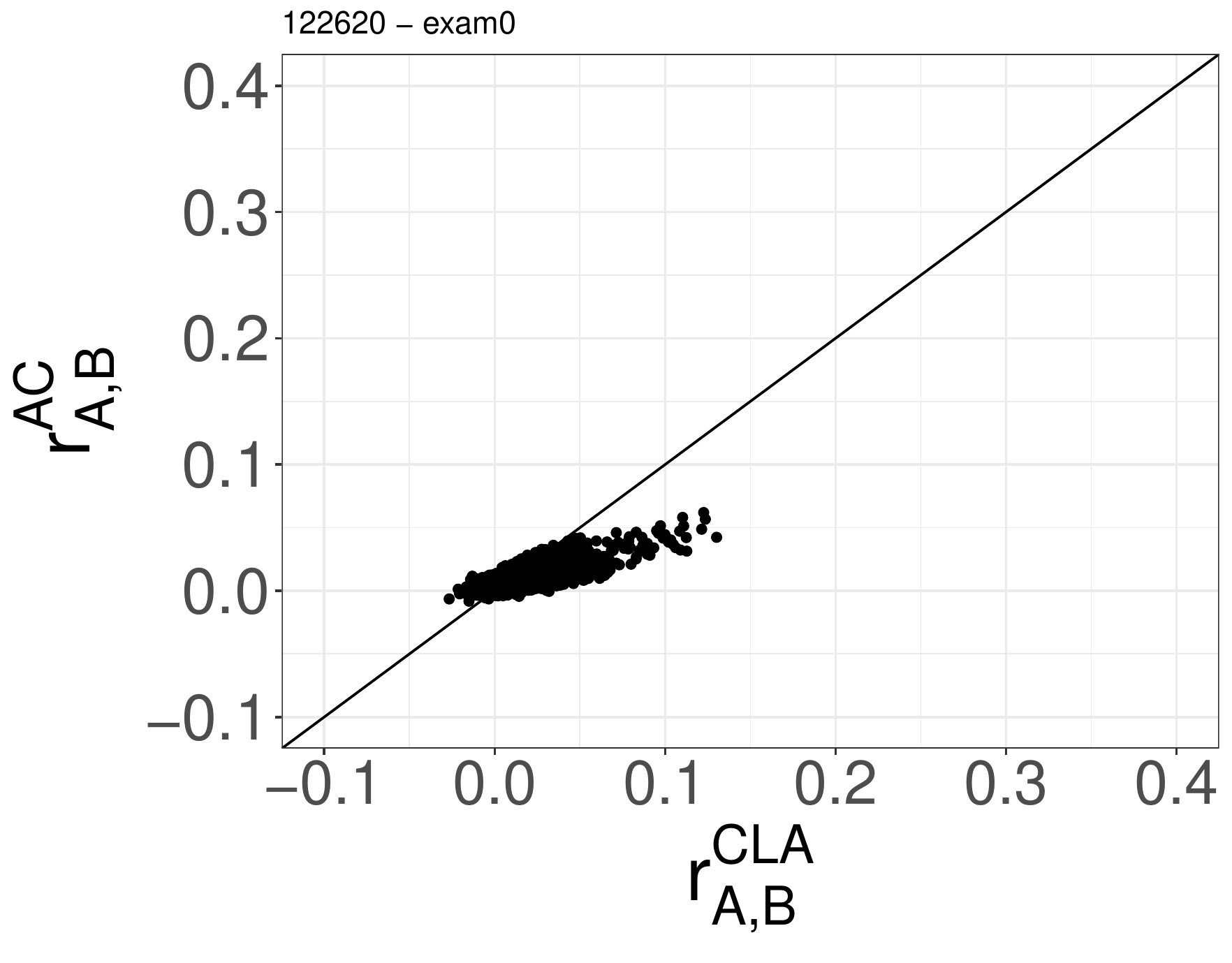} 
			\caption{Subject 2 -- session 0}
	\end{subfigure}
	~
    \begin{subfigure}{0.31\textwidth}
			\centering
			\includegraphics[trim=0cm 0cm 0cm .7cm, clip=true, width=\linewidth]{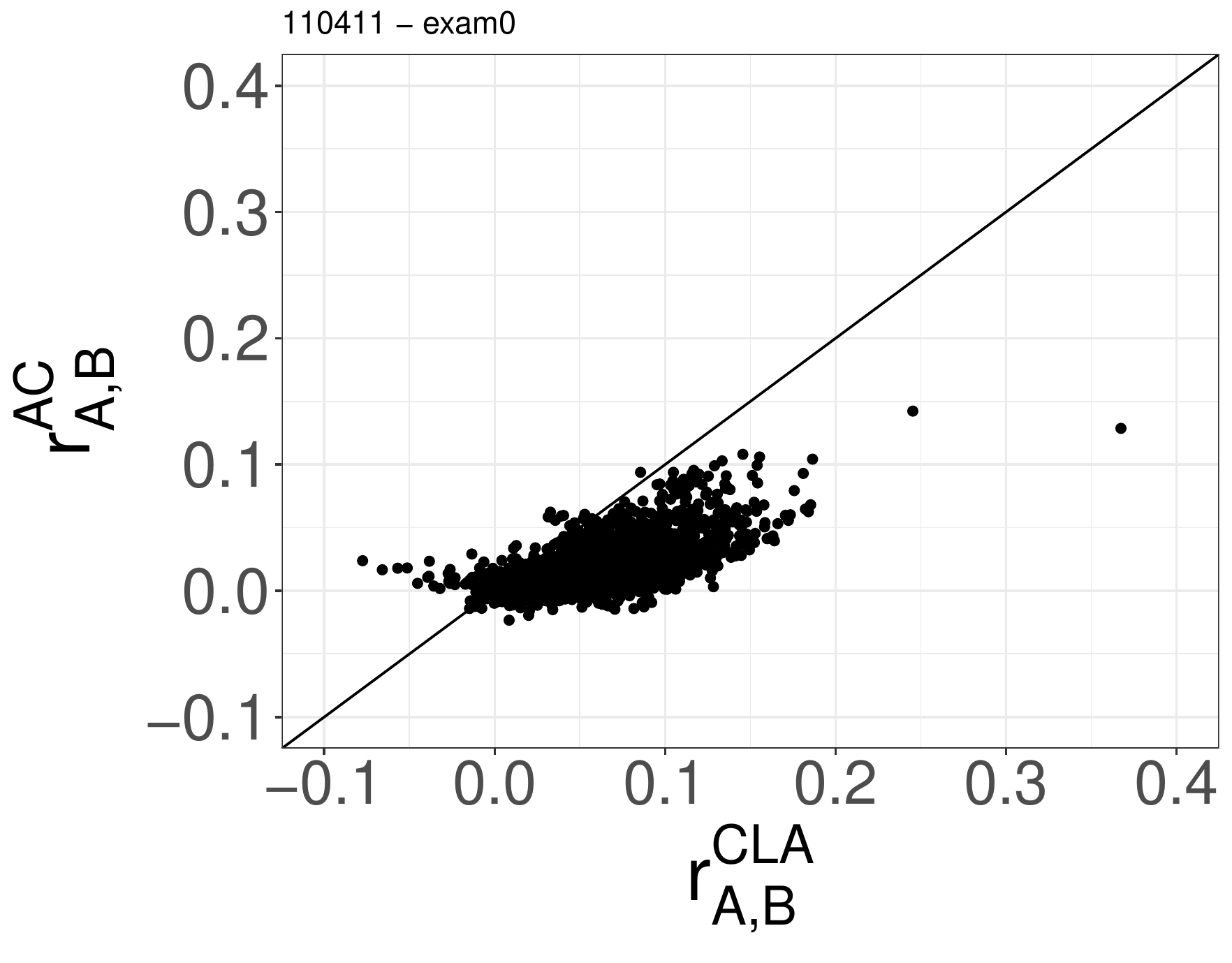} 
			\caption{Subject 3 -- session 0}
	\end{subfigure}
	\caption{
 Inter-correlation coefficients estimated using $r^{AC}_{A,B}$ against our proposed estimator $r^{CLA}_{A,B}$ for three HCP subjects. Each point represents a pair of brain regions.} 
    \label{fig:scatter_HCP_AC}
\end{figure}

We can then compare the sample distribution of our proposed estimator $r^{CLA}_{A,B}$ with that of the standard estimator $r^{CA}_{A,B}$ (cf. Figure \ref{fig:scatter_HCP_CA}) and of $r^{AC}_{A,B}$ (cf. Figure \ref{fig:scatter_HCP_AC}). Overall, and as expected from equations \eqref{eq:vox} and \eqref{eq:rCA} and Corollary \ref{cor:consistency}, the correlation of averages $r^{CA}_{A,B}$ values are higher than that of $r^{CLA}_{A,B}$, while the sample 
values of the average of correlations estimator $r^{AC}_{A,B}$ are lower. In terms of functional connectivity, this means using the $r^{CA}_{A,B}$ estimator could lead to falsely detecting edges, while using $r^{AC}_{A,B}$ could lead to missing edges. These results are in accordance with what was observed in the rat data.


Since we have access to two separate sessions for each subject, we then evaluate the reproducibility of our estimator. To do so, for each subject, we calculate the Concordance Correlation Coefficient (CCC) \citep{lin_concordance_1989} between the inter-correlations estimates from their two sessions. The CCC is scaled between $-1$ and $1$, with $1$ corresponding to a perfect concordance. This means that the higher the CCC, the more reproducible the estimator. The estimator $r^{CLA}_{A,B}$ exhibits the highest CCC, with an average (variance) across the 35 subjects of $0.69$ ($0.03$), while that of $r^{CA}_{A,B}$ is $0.63$ ($0.02$) and $r^{AC}_{A,B}$ is $0.67$ ($0.04$). Our proposed estimator hence improves reproducibility over existing estimators.

\section{Conclusion}
\label{sec:conclusion}
In this paper, we proposed 
a novel and non-parametric estimator of the correlation between groups of arbitrarily dependent variables in the presence of noise. We devised 
a clustering-based approach that simultaneously reduces the impact of noise and intra-correlation through judicious 
aggregation. We then proved that for an appropriate choice of cut-off heights of the dendrograms thus generated, our proposed estimator 
is a consistent estimator of the population inter-correlation. Moreover, our method yields both point estimates and a corresponding sample distribution that could be used, for instance, for uncertainty quantification. We conducted experiments on synthetic data that showed our proposed estimator surpasses popular existing methods in terms of quality, and demonstrated the effectiveness and reproducibility of our approach on real-world datasets.

\paragraph{Supplementary Materials} Proofs of the Theorems are available in the appendix. Discussion about the relaxation of assumptions on the noise, as well as additional details and results on the synthetic datasets are available in the supplementary materials. Source code, including a notebook detailing how to reproduce the figures of this paper, is available at: \url{https://gitlab.inria.fr/q-func/clustcorr}. 

\paragraph{Funding}
This work was supported by the project Q-FunC from Agence Nationale de
la Recherche under grant number ANR-20-NEUC-0003-02 and the NSF grant IIS-2135859.

\appendix
\section{Proof of Theorem \ref{th:ineq.h}}

The proof follows from the properties of hierarchical clustering. In the context of Ward's linkage, the distance between two clusters $\nu_1$ and $\nu_2$ is defined according to \citet[p.~230]{1990ClusteringBook} as: 
\begin{equation}
    \label{eq:ward.dist}
    D(\nu_1, \nu_2) = \sqrt{ \frac{2 \cdot |\nu_1| \cdot |\nu_2| }{|\nu_1| + |\nu_2|} \cdot 
    \left\| \,
    \overline{\textbf{U}}^{\nu_1} - \overline{\textbf{U}}^{\nu_2} 
    \right\|^2 
    },
\end{equation}
where $\overline{\textbf{U}}^{\nu_1}$ is the centroid and $|\nu_1|$ the cardinality of cluster $\nu_1$.
Consider a region $A$ and fix a cut-off height $h_A$. Then, from properties of agglomerative clustering, for any 
cluster $\nu_A$, and for all pairs of U-scores $\textbf{U}_i^{A}, \textbf{U}_{i^{\prime}}^{A}$ inside $\nu_A$, $D( \{ \textbf{U}_i^{A} \} , \{ \textbf{U}_{i^{\prime}}^{A} \}  ) \leq h_A $. Therefore, by combining this inequality with properties of the U-scores \citep{hero2011}, the 
sample intra-correlation can be lower-bounded by a function of $h_A$:
\begin{align}
    1 - \frac{h_A^2}{2} \ \leq \ 1 - \frac{ \| \textbf{U}_i^A - \textbf{U}_{i^{\prime}}^A \|^2 }{2} \ = r_{i,i^{\prime}}^{A,A},
\end{align}
which implies the left-hand side of \eqref{eq:ineq.h}. The right-hand side follows from properties of correlation coefficients. This concludes the proof. 

\section{Proof of Theorem \ref{th:consistent}}
For two clusters $\nu_A, \nu_B$ in regions $A,B$, from \eqref{eq:rCLAdef},
  \begin{align}
        r^{CLA}_{\nu_A, \nu_B} = \frac{ \widehat{Cov}(\, \overline{\textbf{Y}}^{\nu_A}, \overline{\textbf{Y}}^{\nu_B}) }{ \sqrt{\widehat{Var}(\, \overline{\textbf{Y}}^{\nu_A}) \cdot \widehat{Var}(\, \overline{\textbf{Y}}^{\nu_B}) } }.
    \end{align}
Since we have assumed variables are temporally i.i.d., and according to the model definition (cf. Section \ref{sec:prelim}), as $n$ tends towards infinity,
\begin{align}
    \widehat{Cov}(\, \overline{\textbf{Y}}^{\nu_A}, \overline{\textbf{Y}}^{\nu_B}) \stackrel{a.s.}{\to} Cov(\, \overline{Y}^{\nu_A}(t), \overline{Y}^{\nu_B}(t) ),
\end{align} 
for any time point $t$ and where
\begin{align}
    Cov(\, \overline{Y}^{\nu_A}(t), \overline{Y}^{\nu_B}(t) ) & = \frac{1}{|\nu_A| \cdot |\nu_B| } \sum\limits_{i=1}^{|\nu_A|} \sum\limits_{j=1}^{|\nu_B|} Cov(Y_i^A(t), Y_j^B(t)) \nonumber \\
    & = \frac{1}{|\nu_A| \cdot |\nu_B| } \sum\limits_{i=1}^{|\nu_A|} \sum\limits_{j=1}^{|\nu_B|} \sigma_A \sigma_B \rho^{A,B} \nonumber \\ 
    & = \sigma_A \sigma_B \rho^{A,B},
\end{align}
and, from equation \eqref{eq:model},
\begin{align}
    \label{eq:var.as}
    \widehat{Var}(\, \overline{\textbf{Y}}^{\nu_A}) \stackrel{a.s.}{\to} Var(\, \overline{Y}^{\nu_A}(t)) \ = \ \sigma_A^2 \cdot \frac{1}{|\nu_A|^2} \cdot \sum\limits_{i,i^\prime=1}^{|\nu_A|} \eta_{i,i^\prime}^{A} + \frac{\gamma_A^2 }{|\nu_A| },
\end{align}
which gives \eqref{eq:consistency}, and concludes the proof.

\bibliographystyle{Perfect}

\bibliography{biblio_cor_paper}

\begin{thebibliography}{34}
\newcommand{\enquote}[1]{``#1''}
\expandafter\ifx\csname natexlab\endcsname\relax\def\natexlab#1{#1}\fi

\bibitem[\protect\citename{Achard et~al.,
  }2011]{Achard2011fMRIFunctionalConnectivity}
Achard, S., Coeurjolly, J.F., Marcillaud, R., and Richiardi, J. (2011),
  ``{fMRI} functional connectivity estimators robust to region size bias,''
  {\em Proceedings of {IEEE} {W}orkshop on {S}tatistical {S}ignal {P}rocessing
  ({SSP})\/},  813--816.

\bibitem[\protect\citename{Achard et~al., }2020]{Achard2020RobustCF}
Achard, S., Coeurjolly, J.-F., Lafaye~de Micheaux, P., and Richiardi, J.
  (2020), ``Robust correlation for aggregated data with spatial
  characteristics,'' {\em arXiv:2011.08269\/}.

\bibitem[\protect\citename{Achard et~al., }2006]{achard.2006.1}
Achard, S., Salvador, R., Whitcher, B., Suckling, J., and Bullmore, E. (2006),
  ``A Resilient, Low-Frequency, Small-World Human Brain Functional Network with
  Highly Connected Association Cortical Hubs,'' {\em The Journal of
  Neuroscience\/}, 26, 1, 63--72.

\bibitem[\protect\citename{Becq et~al.,
  }2020{\natexlab{a}}]{becq_10.1088/1741-2552/ab9fec}
Becq, G. J.-P.~C., Barbier, E., and Achard, S. (2020{\natexlab{a}}), ``Brain
  networks of rats under anesthesia using resting-state fMRI: comparison with
  dead rats, random noise and generative models of networks{\natexlab{a}},''
  {\em Journal of Neural Engineering\/}{\natexlab{a}}, 17, 045012.

\bibitem[\protect\citename{Becq et~al.,
  }2020{\natexlab{b}}]{guillaume2020functional}
Becq, G. J.-P.~C., Habet, T, Collomb, N., Faucher, M., Delon-Martin, C.,
  Coizet, V., Achard, S., and Barbier, E.~L. (2020{\natexlab{b}}), ``Functional
  connectivity is preserved but reorganized across several anesthetic
  regimes{\natexlab{b}},'' {\em NeuroImage\/}{\natexlab{b}}, 219, 116945.

\bibitem[\protect\citename{Bolt et~al., }2017]{Bolt2017CA}
Bolt, T., Nomi, J.~S., Rubinov, M., and Uddin, L.~Q. (2017), ``Correspondence
  between evoked and intrinsic functional brain network configurations,'' {\em
  Human Brain Mapping\/}, 38, 4, 1992--2007.

\bibitem[\protect\citename{Cameron and Miller, }2015]{ColinCameron2015APG}
Cameron, A.~C. and Miller, D.~L. (2015), ``A Practitioner’s Guide to
  Cluster-Robust Inference,'' {\em The Journal of Human Resources\/}, 50, 317
  -- 372.

\bibitem[\protect\citename{Chavent et~al., }2012]{chavent_clustofvar_2012}
Chavent, M., Kuentz-Simonet, V., Liquet, B., and Saracco, J. (2012),
  ``\textbf{{ClustOfVar}} : {An} \textit{{R}} {Package} for the {Clustering} of
  {Variables},'' {\em Journal of Statistical Software\/}, 50, 13.

\bibitem[\protect\citename{De~Vico~Fallani et~al., }2014]{Fallani2014}
De~Vico~Fallani, F., Richiardi, J., Chavez, M., and Achard, S. (2014), ``Graph
  analysis of functional brain networks: practical issues in translational
  neuroscience,'' {\em Philosophical Transactions of the Royal Society B:
  Biological Sciences\/}, 369, 1653, 20130521.

\bibitem[\protect\citename{Dhillon et~al., }2003]{dhillon_diametrical_2003}
Dhillon, I.~S., Marcotte, E.~M., and Roshan, U. (2003), ``Diametrical
  clustering for identifying anti-correlated gene clusters,'' {\em
  Bioinformatics\/}, 19, 13, 1612--1619.

\bibitem[\protect\citename{Elston, }1975]{Elston1975Cor}
Elston, R.~C. (1975), ``{On the correlation between correlations},'' {\em
  Biometrika\/}, 62, 1, 133--140.

\bibitem[\protect\citename{Ester et~al., }1996]{dbscan1996}
Ester, M., Kriegel, H.-P., Sander, J., and Xu, X. (1996), ``A Density-based
  algorithm for discovering clusters in large spatial databases with noise,''
  {\em Proceedings of the Second International Conference on Knowledge
  Discovery and Data Mining\/},  226–231.

\bibitem[\protect\citename{Glasser et~al., }2013]{glasser_minimal_2013}
Glasser, M.~F., Sotiropoulos, S.~N., Wilson, J.~A., Coalson, T.~S., Fischl, B.,
  Andersson, J.~L., Xu, J., Jbabdi, S., Webster, M., Polimeni, J.~R.,
  Van~Essen, D.~C., and Jenkinson, M. (2013), ``The minimal preprocessing
  pipelines for the {Human} {Connectome} {Project},'' {\em NeuroImage\/}, 80,
  105--124.

\bibitem[\protect\citename{Hartigan and Wong, }1979]{kmeans}
Hartigan, J.~A. and Wong, M.~A. (1979), ``Algorithm AS 136: A K-Means
  Clustering Algorithm,'' {\em Journal of the Royal Statistical Society. Series
  C (Applied Statistics)\/}, 28, 1, 100--108.

\bibitem[\protect\citename{Hero and Rajaratnam, }2011]{hero2011}
Hero, A. and Rajaratnam, B. (2011), ``Large Scale Correlation Screening,'' {\em
  Journal of the American Statistical Association\/}, 106, 1540--1552.

\bibitem[\protect\citename{Kaufman and Rousseeuw, }2005]{1990ClusteringBook}
Kaufman, L. and Rousseeuw, P.~J. (2005), {\em Finding groups in data: an
  introduction to cluster analysis\/}, Wiley series in probability and
  mathematical statistics, Wiley.

\bibitem[\protect\citename{Lin, }1989]{lin_concordance_1989}
Lin, Lawrence I-Kuei (1989), ``A {Concordance} {Correlation} {Coefficient} to
  {Evaluate} {Reproducibility},'' {\em Biometrics\/}, 45, 1, 255--268.

\bibitem[\protect\citename{Lindskog, }2000]{lindskog2000linear}
Lindskog, F. (2000), ``Linear Correlation Estimation,'' {\em Risklab Research
  Paper, ETH-Zentrum, Zürich\/}.

\bibitem[\protect\citename{Liu et~al., }2017]{LiuSmoothingFmri}
Liu, P., Calhoun, V., and Chen, Z. (2017), ``Functional overestimation due to
  spatial smoothing of fMRI data,'' {\em Journal of Neuroscience Methods\/},
  291, 1--12.

\bibitem[\protect\citename{Matzke et~al., }2017]{2017MatzkeBayesCorrError}
Matzke, D., Ly, A., Selker, R., Weeda, W.~D., Scheibehenne, B., Lee, M.~D., and
  Wagenmakers, E.-J. (2017), ``Bayesian Inference for Correlations in the
  Presence of Measurement Error and Estimation Uncertainty,'' {\em Collabra:
  Psychology\/}, 3, 1, 25.

\bibitem[\protect\citename{Murtagh and Legendre, }2014]{Murtagh2014hclust}
Murtagh, F. and Legendre, P. (2014), ``Ward’s Hierarchical Agglomerative
  Clustering Method: Which Algorithms Implement Ward’s Criterion?,'' {\em
  Journal of Classification\/}, 31, 3, 274--295.

\bibitem[\protect\citename{Ogawa, }2021]{Akitoshi2021CA}
Ogawa, A. (2021), ``Time-varying measures of cerebral network centrality
  correlate with visual saliency during movie watching,'' {\em Brain and
  Behavior\/}, 11, 9, e2334.

\bibitem[\protect\citename{Ostroff, }1993]{Ostroff19993aggregated}
Ostroff, C. (1993), ``Comparing Correlations Based on Individual-Level and
  Aggregated Data,'' {\em Journal of Applied Psychology\/}, 78, 569--582.

\bibitem[\protect\citename{Ribeiro and Diggle, }2001]{geoR}
Ribeiro, P.~J. and Diggle, P.~J. (2001), ``geo{R}: a package for geostatistical
  analysis,'' {\em R-NEWS\/}, 1, 2, 14--18.

\bibitem[\protect\citename{Rosner et~al., }1977]{rosner.1977.1}
Rosner, B., Donner, A., and Hennekens, C.~H. (1977), ``Estimation of interclass
  correlation from familial data,'' {\em Applied Statistics\/}, 26, 179--187.

\bibitem[\protect\citename{Saccenti et~al., }2020]{Saccenti2020CorruptionOT}
Saccenti, E., Hendriks, M. M. W.~B., and Smilde, A.~K. (2020), ``Corruption of
  the Pearson correlation coefficient by measurement error and its estimation,
  bias, and correction under different error models,'' {\em Scientific
  Reports\/}, 10, 438.

\bibitem[\protect\citename{Shevlyakov and Smirnov,
  }2016]{Shevlyakov_Smirnov_2016}
Shevlyakov, G. and Smirnov, P. (2016), ``Robust Estimation of the Correlation
  Coefficient: An Attempt of Survey,'' {\em Austrian Journal of Statistics\/},
  40, 147–156.

\bibitem[\protect\citename{Srivastava and Keen, }1988]{Srivastava1988intercorr}
Srivastava, M.~S. and Keen, K.~J. (1988), ``Estimation of the Interclass
  Correlation Coefficient,'' {\em Biometrika\/}, 75, 4, 731--739.

\bibitem[\protect\citename{Termenon et~al., }2016]{termenon2016reliability}
Termenon, M., Jaillard, A., Delon-Martin, C., and Achard, S. (2016),
  ``Reliability of graph analysis of resting state {fMRI} using test-retest
  dataset from the Human Connectome Project,'' {\em NeuroImage\/}, 142,
  172--187.

\bibitem[\protect\citename{Vigneau et~al., }2015]{vigneau_clustvarlv_2015}
Vigneau, E., Chen, M., and Qannari, E.~M. (2015), ``{ClustVarLV}: {An} {R}
  {Package} for the {Clustering} of {Variables} {Around} {Latent}
  {Variables},'' {\em The R Journal\/}, 7, 2, 134.

\bibitem[\protect\citename{Ward, }1963]{Ward1963}
Ward, J.~H. (1963), ``Hierarchical Grouping to Optimize an Objective
  Function,'' {\em Journal of the American Statistical Association\/}, 58, 301,
  236--244.

\bibitem[\protect\citename{Wigley et~al., }1984]{1984wigleyAvgtimeseries}
Wigley, T. M.~L., Briffa, K.~R., and Jones, P.~D. (1984), ``On the Average
  Value of Correlated Time Series, with Applications in Dendroclimatology and
  Hydrometeorology,'' {\em Journal of Applied Meteorology and Climatology\/},
  23, 2, 201--213.

\bibitem[\protect\citename{Wilson, }2010]{2010wilson_families}
Wilson, C. (2010), {\em A study of relationships between family members using
  familial correlations\/}, D.Phil. thesis, Old Dominion University Libraries.

\bibitem[\protect\citename{Zhang et~al., }2016]{Zhang2016CA}
Zhang, C., Cahill, N., Arbabshirani, M., White, T., Baum, S., and Michael, A.
  (2016), ``Sex and Age Effects of Functional Connectivity in Early
  Adulthood,'' {\em Brain Connectivity\/}, 6, 700--713.

\end{thebibliography}

\appendix

\end{document}